\documentclass[12pt]{article}
\usepackage{amsmath}
\usepackage{amssymb}
\usepackage{graphicx,psfrag,epsf}
\usepackage{enumerate}
\usepackage{natbib}
\usepackage{url} % not crucial - just used below for the URL 
\usepackage{color}
\usepackage{amsthm}
\usepackage{amsfonts}
\usepackage[normalem]{ulem}
\usepackage{xcolor}

\newcommand\bluesout{\bgroup\markoverwith{\textcolor{blue}{\rule[0.5ex]{2pt}{0.4pt}}}\ULon}

\DeclareMathOperator{\Bin}{Bin}

\newcommand{\Real}{\mathbb{R}}

\DeclareMathOperator{\sign}{sign}

\DeclareMathOperator{\As}{\mathcal{A}}

\newcommand{\argmin}{\qopname\relax m{arg\,min}}

\def\bfred#1{{\color{red}\bf#1}}

\newcommand{\vbeta}{\mbox{\boldmath $\beta$}}
\newcommand{\vtheta}{\mbox{\boldmath $\theta$}}
\newcommand{\vmu}{\mbox{\boldmath $\mu$}}
\newcommand{\veta}{\mbox{\boldmath $\eta$}}
\newcommand{\vSigma}{\mbox{\boldmath $\Sigma$}}
\newcommand{\vepsilon}{\mbox{\boldmath $\epsilon$}}
\newcommand{\vDelta}{\mbox{\boldmath $\Delta$}}
\newcommand{\vomega}{\mbox{\boldmath $\omega$}}

\newcommand{\va}{{\bf a}}
\newcommand{\vb}{{\bf b}}

\newcommand{\ve}{{\bf e}}

\newcommand{\vW}{{\bf W}}

\newcommand{\vE}{{\bf E}}

\newcommand{\vs}{{\bf s}}
\newcommand{\vu}{{\bf u}}

\newcommand{\vx}{{\bf x}}
\newcommand{\vy}{{\bf y}}

\newcommand{\vA}{{\bf A}}
\newcommand{\vB}{{\bf B}}
\newcommand{\vC}{{\bf C}}

\newcommand{\vI}{{\bf I}}

\newcommand{\vL}{{\bf L}}

\newcommand{\vX}{{\bf X}}
\newcommand{\vY}{{\bf Y}}
\newcommand{\vZ}{{\bf Z}}
\newcommand{\vD}{{\bf D}}

\def\bfred#1{{\color{red}\bf#1}}

\newtheorem{theorem}{Theorem}
\newtheorem{corollary}{Corollary}
\newtheorem{condition}{Condition}

\newcommand{\blind}{1}

\begin{document}

\def\spacingset#1{\renewcommand{\baselinestretch}%
{#1}\small\normalsize} \spacingset{1}

%%%%%%%%%%%%%%%%%%%%%%%%%%%%%%%%%%%%%%%%%%%%%%%%%%%%%%%%%%%%%%%%%%%%%%%%%%%%%%

\if1\blind
{
  \title{\bf On the Use of Minimum Penalties in Statistical Learning}
  \author{Ben Sherwood \thanks{
    The authors gratefully acknowledge the support of the West Virginia Clinical and Translational Science Institute (NIH/NIGMS Award Number U54GM104942)  and note that this work is partially supported by  by the  National Science Foundation (NSF) Major 
   Research Instrumentation Program (MRI Award 1726534) and the John Chambers College of Business and Economics Summer Research Program.}\hspace{.2cm}\\
    School of Business, University of Kansas\\
    and \\
    Bradley S. Price \\
    John Chambers College of Business and Economics,  West Virginia University}
  \maketitle
} \fi

\if0\blind
{
  \bigskip
  \bigskip
  \bigskip
  \begin{center}
    {\LARGE\bf A Minimum Penalty for Multivariate Regression}
\end{center}
  \medskip
} \fi

\bigskip
\begin{abstract}
Modern multivariate machine learning and statistical methodologies estimate parameters of interest 
while leveraging prior knowledge of the association between outcome variables.  The methods that do allow for estimation of relationships do so typically through an error covariance matrix in multivariate regression which does not scale to other types of models.  In this article we proposed the MinPEN framework  to simultaneously estimate regression coefficients associated with the multivariate regression model and the relationships between outcome variables using mild assumptions.  The MinPen framework utilizes a novel penalty based on the minimum function to exploit detected relationships between responses.  An iterative algorithm that generalizes current state of the art methods is proposed as a solution to the non-convex optimization that is required to obtain estimates.  Theoretical results such as high dimensional convergence rates, model selection consistency, and a framework for post selection inference are provided.  We extend the proposed MinPen framework to other exponential family loss functions, with a specific focus on multiple binomial responses.  Tuning parameter selection is also addressed.  Finally, simulations and two data examples are presented to show the finite sample properties of this framework.

\end{abstract}

\noindent%
\vfill

\newpage
\spacingset{1.45} % DON'T change the spacing!
\section{Introduction}
\label{sec:intro}

Multivariate, also known as multiple response or multi-task, regression jointly models multiple responses (outcomes) given a common set of predictors (inputs). Joint modeling of responses,  as opposed to modeling each response separately,  is only interesting if the responses are related. Leveraging these relationships should improve estimation of the model,  but in practice these relationships may not always be known.  In this work we propose a framework to simultaneously estimate the regression coefficients and detect relationships between the response variables. The framework also allows for simultaneous variable selection.  

%Multivariate regression is widely used in many fields including finance, genetics, and advanced manufacturing. Also known as multiple response regression or multi-task regression the goal is to jointly model multiple outcomes give a uniform set of inputs.  In this manuscript we propose a new framework for estimating 
%the regression coefficients associated with the multivariate regression problem while also providing a framework to detect novel relationships between mean functions of the  response (outcome) variables of interest.  
Consider the sample of independent data, $\{\vx_i,\vy_i\}_{i=1}^n$, where $\vx_i \in \Real^p$ and $\vy_i \in \Real^r$ with a model of 
\begin{equation}
\label{lin_mod} 
\vy_i = \vB^{*T}\vx_i + \vepsilon_i,
\end{equation}
where $\vB^*=(\vbeta_1^*,\ldots,\vbeta_r^*) \in \Real^{p \times r}$ and $\vepsilon_i \in \Real^r$ are i.i.d. random vectors with mean zero and covariance matrix $\vSigma$. A popular estimator for $\vB^*$ is 

\begin{equation}
\label{OLS}
\widehat{\vB}=\argmin_{\vbeta_1,\ldots,\vbeta_r \in \Real^p} \frac{1}{2n}\sum_{k=1}^r\sum_{i=1}^n (y_{ik}-\vx_i^T\vbeta_k)^2,
\end{equation}
which has desirable qualities such as consistency and asymptotic normality under mild conditions, and is the maximum likelihood estimator if the errors are from a multivariate normal distribution. However, there is something unsatisfying about not using correlation of the responses in the estimation. To improve efficiency in estimation of coefficients, while accounting for correlation across the errors \citet{rothman2010} (MRCE), \citet{witten09} (SCOUT) and \citet{lee2012} have all proposed methods which simultaneously estimates the regression coefficients and the inverse covariance matrix of the errors. These methods all aim to improve estimation of regression coefficients by exploiting correlation across the responses that is not explained by the predictors.  

%\bfred{(BP: replacing relationships and association instead of correlation}.
We propose an alternative approach that focuses on relationships across the responses that can be explained, due to similarity in the regression coefficients.  
Specifically, we consider the kth and lth response related if $\vbeta_k^*$ is similar to $\vbeta_l^*$ or $-\vbeta_l^*$.  If relationships were known \emph{a priori}, then for each response $k$ one could define three disjoint sets: (1) $P_k$, a set of responses positively related to response $k$; (2) $N_k$, a set of responses negatively related to response $k$; and (3) $Z_k$ a set of responses that are not related to $k$. 
%the disjoint sets $P_k$, $N_k$ and $Z_k$. %Where $P_k\cup N_k \cup Z_k= \{1,\ldots, r\}\setminus \{k\}$, 
%Where $P_k$ defines the responses that are positively related to response $k$, $N_k$ contains the set of responses that are negatively related to response $k$, and $Z_k$ contains the set of responses that are not related to response $k$.  
These relationships could be used to improve estimation using a penalized likelihood such as

\begin{equation}
\label{setup}
\argmin_{\vB \in \Real ^{p \times r}} \frac{1}{2n}\sum_{k=1}^r (\vy_k-\vx_i^T\vbeta_k)^T(\vy_k-\vx_i^T\vbeta_k) +\frac{\gamma}{2}\sum_{k=1}^r\left(\sum_{l \in P_k} \|\vbeta_l-\vbeta_k\|_2^2 + \sum_{m \in N_k} \|\vbeta_m +\vbeta_k\|_2^2 +\sum_{s \in Z_k} \|\vbeta_k\|_2^2\right) ,
\end{equation} 
where, for a vector $a\in \Real^d$ define $\|a\|_q=(\sum_{j=1}^d a_{j}^q)^{\frac{1}{q}}$ as the $L_q$ norm and $\gamma>0$ is a tuning parameter to promote similarities between groups of related coefficients. 

The method we propose is a generalization of \eqref{setup} when the sets $P_k$, $N_k$ and $Z_k$ are not known \emph{a priori} and uses an $L_1$ penalty for feature selection. The proposed method simultaneously estimates these unknown sets, while using the framework of \eqref{setup} to include these detected relationships in the estimation. %, a property that maintains in cases where the response may not be continuous such as generalized linear models. 
To simultaneously estimate the group structure we propose a novel penalty based on the minimum function. %The resulting penalized objective function is continuous, but non-convex. 
The penalized objective function is non-convex but can be re-stated as solving a fixed number of convex problems. However, this quickly becomes computationally intractable for even small values of $r$. Unlike the earlier cited work which focuses on structure due to relationships in the errors, this approach can be easily generalizable to simultaneous modeling of binary response variables or other non-continuous values by replacing the least squares loss function with the appropriate generalized linear model loss function.  

%\bluesout{\bfred{The algorithm proposed to obtain estimates} allows for simultaneous coefficient estimation, variable selection and grouping of the responses \bfred{which generalizes state of the art methods in the literature} \bfred{and can be easily extended to work with other loss functions. } }  %This algorithm can also be extended to any generalized linear model setting and we demonstrate how to extend it to logistic regression. 

In the least squares case we provide rates of convergence, model selection consistency and post selection inference results which hold for high-dimensional predictors, $p>>n$. The results hold for any grouping of the response variables and thus do not depend on the relationships between responses to be estimated correctly.
%correct grouping structure being found. %The commonly used framework presented in \citet{negahban2012} to derive convergence rates for penalized estimators does not directly apply to the proposed method due to the ridge fusion penalty. 
To derive rates of convergence under standard conditions we generalize the results of \citet{negahban2012}, which do not directly apply because of the ridge fusion penalty used in this framework.

Our proposal builds on methods which simultaneously estimate clusters and univariate regression models \citep{witten14}, multivariate regression models \citep{price_sherwood}, and precision matrices \citep{price_molstad_sherwood}. All of which propose iterative algorithms that alternate between estimating the clustering structure, using k-means, and the model of interest. These approaches only accommodate positive relationships and rely on k-means, which provides distinct sets of clusters and may be unreliable for high-dimensional problems. The proposed method allows for more complex structures in the response variables and provides a global minimum at each iteration. %\bfblue{ some remark on how our algorithm may be more stable than k-means}. 

{Extensive work has been done on exploiting relationships in multivariate regression to improve efficiency. %What follows is an inevitably incomplete list of some major contributions to the field. 
For a recent survey of the field see \cite{pas_review} and citations within.   Reduced rank regression reduces the dimension of the problem by constraining the rank of the coefficient matrix $\vB$ while minimizing the multivariate least squares objective function to find a set of latent variables that increase prediction accuracy \citep{anderson1951, velu2013, chen2016sparse}. Various group lasso penalties have been used to promote structure across response or a combination of response and predictors. \citep{kim2012tree,li2015} Instead of the proposed approach of  simultaneously estimating relationships and coefficients, many have used a two-stage clustering approach. In the first stage the clusters are estimated and in the second stage the structure is incorporated into the estimation of the coefficients. Examples include hierarchical clustering for tree guided lasso \citep{kim2012tree}, convex clustering in conjoint analysis \citep{chen_2016}, separating global and response specific features \citep{xu2015exploiting} and replacing high dimensional responses or predictors with clusters \citep{buhlmanCluster, bird}. 

Other methods in univariate regression,  such as the GRACE method, utilize known relationships, or relationships that are estimated \emph{a priori},  to assist in increasing accuracy and select relevant variables \citep{li_08, li_10}.  Others such as \citet{zhao_16} have investigated statistical inference frameworks for univariate graph constrained models, that also accommodate error in the estimation of the graph Laplacian.

The key difference between the proposed method and state of the art methods in the literature, is that our method simultaneously detects relationships between the responses, regardless of sign, and estimates regression coefficients without needing to estimate a covariance matrix of the errors. In Section \ref{Method} we introduce the MinPen framework with the least squares loss function along with theoretical results and the proposed algorithm. In Section \ref{bin_res} we extend the framework to multiple binomial responses with discussion of how it may generalized to other exponential family based loss functions.  Sections \ref{olssim} and 5 present simulations to investigate finite sample properties of the proposed methods.  Finally we present examples of our methodology in applications in genomics and substance abuse overdoses in Section \ref{applications}.    

\section{Least Squares Model}
\label{Method}
\subsection{Method}
\label{norm_method}

We consider estimating $\vB^*$ in \eqref{lin_mod} when there may be similarities in the mean functions of the different responses. Define,
\begin{equation}
\label{minpen0}
P(\vbeta_l,\vbeta_k)=\min(\|\vbeta_l-\vbeta_k\|_2^2,\|\vbeta_l+\vbeta_k\|_2^2,\|\vbeta_l\|_2^2).
\end{equation}

If $P(\vbeta^*_l,\vbeta^*_k) = \|\vbeta_l^*-\vbeta^*_k\|_2^2$ this implies that the mean functions for response $k$ and $l$ are positively related. If $P(\vbeta^*_l,\vbeta^*_k) = \|\vbeta_l^*+\vbeta_k^*\|_2^2$ this implies negative relationship between the two mean functions, while $P(\vbeta^*_l,\vbeta^*_k)=\|\vbeta^*_l\|_2^2$ implies little or no relationship between the mean functions. Therefore a penalty based on the minimum function can be used to simultaneously identify and leverage relationships between mean functions. Motivated by this, we propose the  minimum penalty elastic net multivariate regression (MinPen) estimator as the solution to
\begin{equation}
\label{minpen1}
\argmin_{\vB \in \Real^{p\times r}} \frac{1}{2n}\sum_{i=1}^n\sum_{k=1}^r (y_{ik}-\vx^T_i\vbeta_k)^2 +\delta\sum_{k=1}^r \|\vbeta_j\|_1 +\frac{\gamma}{2}\sum_{l=1}^{r}\sum_{k=1, k\neq l}^r P(\vbeta_l,\vbeta_k),
\end{equation}
where $\delta$ and $\gamma$ are non-negative tuning parameters specified by the user. We refer to this as an elastic net estimator as it uses the combination of the lasso and ridge penalties as first introduced by \cite{zou05}.  The lasso penalty, associated with tuning parameter $\delta$, is used to simultaneously perform variable selection and estimate regression coefficients.   While, the proposed ridge type penalty, associated with $\gamma$, is a non-convex penalty that simultaneously identifies and exploits relationships between coefficient vectors of different responses.

%We introduce a ridge type \emph{minimum penalty} which we propose here, associated with tuning parameter $\gamma$, which is a non-convex penalty used to exploit similarities in regression coefficient vectors if they exist.  This penalty detects whether the coefficient vectors have a positive association, negative association, or no association, and performs the appropriate shrinkage. %We note that other penalties could be used inside the minimum function in the penalty associated with $\gamma$, which we refer to generally as a \emph{minimum penalty}.  We do note others such as 

%\cite{luo2012two} \bluesout{proposed using a penalty for network predictors in univariate regression that at it's limit is a maximum function, but to our knowledge we are the first to investigate the use of a minimum penalty,  specifically in the multivariate setting.  While this formulation of the model gives a concise definition of the estimator, there are other formulation of the problem that give some other insight into the methodological impacts this estimator can have.}

Both our theory and algorithms will deal with a vectorized version of the solution to \eqref{minpen1}. For a matrix $\vA$, define $\mbox{vec}(\vA)$ as the vector formed by stacking the columns of $\vA$ on top of one another. Define $\vY=(\vy_1,\ldots,\vy_n)^T \in \Real^{n\times r}$, $\vX=(\vx_1,\ldots,\vx_n)^T \in \Real^{n\times r}$, $\vE=(\epsilon_1,\ldots,\epsilon_n)^T \in \Real^{n \times r}$, $\tilde{\vy} = \mbox{vec}(\vY) \in \Real^{nr}$, $\tilde{\vX} = \vI_r \otimes \vX \in \Real^{nr \times pr}$, $\vbeta^* = \mbox{vec}(\vB^*) \in \Real^{pr}$, $\tilde{\vepsilon} = \mbox{vec}(\vE) \in \Real^{nr}$ and $\tilde{\vSigma} = \vSigma \otimes \vI_n \in \Real^{nr \times nr}$.   Then the vectorized version of \eqref{lin_mod} is
\begin{equation}
\label{vecm_model}
\tilde{\vy} = \tilde{\vX}\vbeta^* + \tilde{\vepsilon}.
\end{equation} 
%Where the covariance of $\tilde{\vepsilon}$ is $\tilde{\vSigma} = \vSigma \otimes \vI_n \in \Real^{nr \times nr}$. 

Define the set 
$$\As = \left\{ \vA \in \{-1,0,1\}^{r(r-1)p \times rp} \middle| \,  ||\vA \vbeta||_2^2 = \sum_{l=1}^{r} \sum_{m=1, l \neq m}^r ||\vbeta_l-d_{lm}\vbeta_m||_2^2 \mbox{ where } d_{lm} \in \{-1,0,1\}\right\}.$$ 
The estimator in \eqref{minpen1} is equivalent to  
\begin{equation}
\label{minpen2c}
%\label{minpen2a}
(\hat{\vbeta},\hat{\vA}) %&=& \argmin_{\vbeta \in \Real^{pr}, \vA \in \As} \frac{1}{2n}\sum_{i=1}^n\sum_{k=1}^r (y_i-\vx^T_i\vbeta_k)^2 +\delta\sum_{k=1}^r \|\vbeta_j\|_1 +\frac{\gamma}{2n}\vbeta^TA^TA\vbeta \\
%\label{minpen2b}
%&=& \underset{\vbeta \in \Real^{pr}}{\mbox{argmin}} \frac{1}{2n} ||\tilde{\vY}-\tilde{\vX}\vbeta||_2^2 + \delta ||\vbeta||_1 + \frac{\gamma}{2} \underset{\vA \in \As}{\min}\left(||\vA\vbeta||_2^2\right) \\
= \underset{\vbeta \in \Real^{pr}}{\mbox{argmin}} \, \underset{\vA \in \As}{\min} \left(\frac{1}{2n} ||\tilde{\vY}-\tilde{\vX}\vbeta||_2^2 + \delta ||\vbeta||_1 + \frac{\gamma}{2} ||\vA\vbeta||_2^2\right).
\end{equation}
Due to the minimum function the objective function is non-convex, which typically makes theoretical study of the global minimizer challenging. However, the global solution to \eqref{minpen2c} is the pair $(\hat{\vbeta},\hat{\vA})$ that minimizes across the $|\As|$ potential convex objective functions. %, all possible combinations of relationships of the $r$ responses.  
So while the objective function is nonconvex, the solution $\hat{\vbeta}$ minimizes a convex function. If the matrix $\vA$ is known \emph{a priori} the estimator of $\vbeta^*$ is
\begin{equation}
\label{grace}
\hat{\vbeta}(\vA)=\argmin_{\vbeta \in \Real^{pr}} \frac{1}{2n}\sum_{i=1}^n\sum_{k=1}^r (y_i-\vx^T_i\vbeta_k)^2 +\delta\sum_{k=1}^r \|\vbeta_k\|_1 +\frac{\gamma}{2n}\vbeta^T\vA^T\vA\vbeta. 
\end{equation}

For our theoretical results we first analyze the estimator $\hat{\vbeta}(\vA)$. Then, using that the solution is one of the potential $|\As|$ solutions, we provide rates of convergence for the global solution, $\hat{\vbeta}$.  {It may be desirable to replace $\As$ with some subset, say $\As_1 \subset \As$. If it is known that two responses are positively related then it does not makes sense to consider the negative or no relationship for those two responses and thus a suitable $\As_1$ could be used instead of $\As$. For this paper we consider the general case, but the results and algorithm presented here will hold if a smaller set is considered.

%\bluesout{which has a similar form to a Grace estimator} \citep{li_08,li_10}. \bluesout{There are two key differences, first the current literature on Grace estimators does not consider multivariate responses and second it assumes $A$ is known. Note that under this formulation it is clear the if we view the MinPen estimator as a graph constrained model, the estimator must simultaneously estimate the regression coefficients and the graph (or matrix $A$).}

%\bluesout{Both formulations of the objective function are useful for discussion of intuition, theoretical discussions, and computational foundations. We find that \eqref{minpen1} is the most intuitive formulation of the problem, we find that the formulation of the objective function in \eqref{minpen2a} is the most useful for developing the theoretical foundations of this work. Finally, we will use a third formulation similar to the optimization show in given by \eqref{setup} to provide an iterative algorithm for obtaining the estimates.}

\subsection{Theoretical Results}
\label{theory}
Mild or standard conditions were used to derive the rates of convergence. Before presenting the conditions, we define some notation that is used in our theorems and conditions. The subspace for the active predictors is defined as
$
\mathcal{M} \equiv \{ \va \in \Real^{pr} | a_j = 0 \mbox{ if } \vbeta^*_j = 0 \},
$ with cardinality $s=|\mathcal{M}|$. The parameter space is separated using projections of vectors into orthogonal complements. Define $\vu_{\mathcal{M}}$ as a projection of a vector $\vu$ into space $\mathcal{M}$, $\mathcal{M}^\perp$ as the orthogonal complement of $\mathcal{M}$ and $\mathcal{C} \equiv \{ \va \in \Real^{pr} \vert \, ||\va_{\mathcal{M}^\perp}||_1 \leq 3||\va_{\mathcal{M}}||_1 \}$. % as  
%$$
%\vu_{\mathcal{M}} \equiv \underset{\vv \in \mathcal{M}}{\mbox{arg min}} ||\vu-\vv||_2.
%$$
%The orthogonal complement of space $\mathcal{M} \subseteq \Real^{p}$ is 
%$$
%\mathcal{M}^\perp \equiv \{ \vv \in \Real^{pr} | \langle \vu, \vv \rangle =0 \mbox{ for all } \vu \in \mathcal{M}(S) \}.
%$$
%In other words, 
%\begin{equation*}
%\mathcal{M}^\perp \equiv \{ \vbeta \in \Real^{pr} | \beta_j = 0 \mbox{ if } \vbeta^*_j \neq 0 \}.
%\end{equation*}
%Similar to \citet{negahban2012}, and many others, some of our conditions reference the following cone
%\begin{equation*}
%\mathcal{C} \equiv \{ \va \in \Real^{pr} \vert \, ||\va_{\mathcal{M}^\perp}||_1 \leq 3||\va_{\mathcal{M}}||_1 \}.
%\end{equation*}
%Rates of convergence were derived using the following conditions. 

\begin{condition}
\label{xscale}
Define $\vx_j \in \Real^p$ to be the $j$th column vector of $\vX$. For all $j \in \{1,\ldots,p\}$ $\frac{||\vx_j||_2^2}{n}\leq 1$.
\end{condition}

\begin{condition}
\label{rec}
There exists positive constants $\kappa_l$ and $\kappa_u$ such that for all $\vtheta \in \mathcal{C}$ that
$
\kappa_l \leq \vtheta^T \frac{1}{n} \tilde{\vX}^T \tilde{\vX} \vtheta \leq \kappa_u.
$
\end{condition}

\begin{condition}
\label{con_error}
The error vector $\vepsilon_j$ has a mean of zero and sub-Gaussian tails for all $j \in \{1,\ldots,r\}$. That is, for all $j \in \{1,\ldots,r\}$ there exists a constant $\sigma_j$ such that for any $\va \in \Real^n$, with $||\va||_2 = 1$, 
\begin{equation*}
P(|\langle \vepsilon_j, \va \rangle| > t) \leq 2 \mbox{exp} \left( - \frac{t^2}{2\sigma_j^2} \right).
\end{equation*}
In addition, there exists a positive constant $\sigma$ such that $\sigma \geq \underset{j}{\max} \, \sigma_j$. 
\end{condition}

\begin{condition}
\label{b_inf}
There is at least one non-zero entry in $\vbeta^*$. 
\end{condition}

Conditions \ref{xscale}-\ref{con_error} are commonly made for penalized estimators, see \citet{negahban2012} and the citations within. The framework in \citet{negahban2012} does not directly apply to \eqref{minpen2c}, because they assume the penalty is a norm which is not the case for our penalty, even for $\vA$ fixed, because of the inclusion of the Ridge penalty.  However, under these conditions we can extend the approach of \citet{negahban2012} and prove rates of convergence under milder conditions then some similar methods \citep{price_sherwood, li_10}. %\bfblue{Grace?}
In our results the upper bound of the tuning parameter $\gamma$ depends on the inverse of $||\vbeta^*||_\infty$, which is well defined under Condition \ref{b_inf}.%, \bfblue{which holds if there is any association between the predictors and responses.} %\bfred{\sout {In practice we expect this condition to typically hold if any thought was put in to the decision about which variables to collect.}}

%The following condition guarantees that at least one of the predictors has a non-zero coefficient. In our proofs the rate of $\gamma$ will have an inverse relationship with the infinity norm of $\vbeta^*$. This condition guarantees that relationship is well defined. Let $s$ be the total number of nonzero coefficients in $\vbeta^*$, the following condition guarantees that $s>0$.

%Without loss of generality assume the first $q$ predictors are non-spares and have at least one non-zero value in the corresponding row of $\vB^*$ and that the remaining $p-q$ predictors are sparse across all responses. In the special case that the all responses have the same sparsity structure then $s=rq$. The following condition guarantees that $\vbeta^*(A) \in \mathcal{M}$, that is the sparsity structure of $\beta^*(A)$ is the same as the sparsity structure of $\vbeta^*$. Let $\vX_{Q} \in \Real^{n \times q}$ be the design matrix of the active predictors and $\vX_{N} \in \Real^{n \times p-q}$ be the design matrix of inactive predictors. Define $\vS_Q = \frac{1}{n} \vX_{Q}^T\vX_{Q}$ and $\vS_N = \frac{1}{n} \vX_N^T\vX_n$. 

\begin{theorem}
\label{thm1_ext}
Assume Conditions \ref{xscale}-\ref{b_inf} hold and $\gamma < \delta/(16r||\vbeta^*||_\infty)$ then for any $\vA \in \As$
\begin{equation*}
P\left[||\hat{\vbeta}(\vA)-\vbeta^*||_2^2 \leq  \frac{9\delta^2}{4\kappa_l^2}s\right] \geq 1 - 2 \mbox{exp} \left[ \frac{-n\delta^2}{32\sigma^2}+\log(rp)\right].
\end{equation*}
%with probability at least $1 - 2 \mbox{exp} \left[ \frac{-n\delta^2}{32\sigma^2}+\log(rp)\right]$. 
\end{theorem}

Proof of Theorem \ref{thm1_ext}, and all other theoretical results, are provided in the supplemental material. Theorem \ref{thm1_ext} can be used to derive convergence rates for $\hat{\vbeta}(\vA)$ for a set sequence of $\delta$. 

\begin{corollary}
\label{thm_c}
Assume Conditions \ref{xscale}-\ref{con_error} hold, $\gamma < \delta/(16r||\vbeta^*||_\infty)$ and $\delta = \sqrt{\frac{64\sigma^2\log(rp)}{n}}$ then for any $\vA \in \As$
\begin{equation*}
P\left[ ||\hat{\vbeta}(\vA)-\vbeta^*||_2^2 \leq 144 \frac{\sigma^2\log(rp)}{n\kappa_l^2}s\right] \geq 1 - 2 \mbox{exp}[-\log(rp)].
\end{equation*}
%with probability at least $1 - 2 \mbox{exp}[-\log(rp)]$. 
\end{corollary}
%\begin{proof}
%Apply Theorem \ref{thm1_ext} with the given rate of $\delta$. 
%\end{proof}

Corollary \ref{thm_c} provides that for fixed $\vA$ and $rp \rightarrow \infty$ that the estimator $\hat{\vbeta}(\vA)$ achieves the same rate of converge as the lasso estimator \citep{negahban2012} and elastic net \citep{smoothLasso}. Using Corollary \ref{thm_c} and noting that $\hat{\vbeta}$ is equivalent to a $\hat{\vbeta}(\hat{\vA})$ we can derive a rate of convergence for $\hat{\vbeta}$, the global minimum of a nonconvex objective function.%, by taking a union across $\As$.

\begin{corollary}
\label{globalRate}
Assume Conditions \ref{xscale}-\ref{con_error} hold, $\gamma < \delta/(16r||\vbeta^*||_\infty)$ and $\delta = \sqrt{\frac{64\sigma^2\log(rp)}{n}}$ then 
\begin{equation*}
P\left[||\hat{\vbeta}-\vbeta^*||_2^2 \leq 144 \frac{\sigma^2\log(rp)}{n\kappa_l^2}s\right] \geq 1 - 2 \mbox{exp}[-\log(rp)+\log(|\As|)].
\end{equation*}
%with probability at least $1 - 2 \mbox{exp}[-\log(rp)+\log(|\As|)]$. 
\end{corollary}

\subsection{Model Selection Consistency}

Let $\dot{\vX}_{(1)} \in \Real^{nr \times s}$ be the submatrix of $\tilde{\vX}$ consisting of the active predictors of the regression defined in \eqref{vecm_model}, and $\dot{\vX}_{(2)} \in \Real^{nr \times pr-s}$ be the submatrix of the remaining predictors. Define $\dot{\vX} = ( \dot{\vX}_{(1)}, \dot{\vX}_{(2)})$ and let $\dot{\vbeta}^*$ and $\dot{\vbeta}(\vA)$ be the similarly re-arranged versions of $\vbeta^*$ and $\hat{\vbeta}(\vA)$, respectively. In addition, let $\dot{\vbeta}^*_{(1)}$ be the first $s$ entries of $\dot{\vbeta}^*$. Note, that we still have
$$
\tilde{y}=\dot{\vX}\dot{\vbeta}^*+\tilde{\vepsilon} = \dot{\vX}_{(1)}\dot{\vbeta}^*_{(1)} + \tilde{\vepsilon}.
$$
Thus model selection consistency results found for $\dot{\vbeta}(\vA)$, also hold for $\hat{\vbeta}(\vA)$.
%\bfred{Assume that we can reformat $\tilde{\vX}$ into the partition} $\dot{\vX}=\left( \dot{\vX}_{(1)}, \dot{\vX}_{(2)}\right)$ such that $\dot{\vX}_{(1)} \in \Real^{nr \times s}$ contains the active predictors of the regression defined in \eqref{vecm_model}, and $\dot{\vX}_{(2)} \in \Real^{nr \times pr-s}$ contains the predictors that are not in the active set. We then define the formulation 
%$$
%\tilde{y}=\dot{\vX}\dot{\vbeta}^*+\tilde{\epsilon}
%$$
%and $\dot{\vbeta}=\dot{\vbeta}(\vA)$ be the corresponding MinPen estimator of $\dot{\beta}^*$.  
% \bfblue{I think this partitioning introduces some potential confusion. For a univariate response it is not a problem because you can sort the data that way without a loss of generality. I suppose you could do that for our problem, but then the definition of everything including $\vbeta$ gets much more complex as $\vbeta^*$ and $\hat{\beta}$ are not originally defined in the format you are suggesting here. This is why I used some more cumbersome notation in the next section. I think we should discuss this at a future meeting. One solution would be to introduce some ``scrambled'' versions of $X$ and $\beta$ say $\bar{X}$ and $\bar{\beta}$ remark on how they are equivalent to the other ones, but for ease of proof we do theory on the scrambled versions here. Alternatively, we could first define $X$ in the ``scrambled form''.}

Define $\dot{\vA}$ such that it relates to $\dot{\vX}$ the same way $\vA$ corresponds to $\vX$. The signed graph Laplacian of $\dot{\vA}  \in \As$ is  %$L$, as \bfblue{(Do we care that $\vA$ needs to be re-arranged? I'm leaning towards no.)}
$$
\vL=\dot{\vA}^T\dot{\vA}=\left(\begin{array}{cc}
\vL_{11} & \vL_{12} \\ 
\vL_{21} & \vL_{22}
\end{array} \right),
$$
where $\vL_{11}$ and $\vL_{22}$ are the signed graph Laplacians of the active and inactive sets respectfully. %the the relationships between variables in the active set which in our case is the relationship between variables \bfred{ in$\dot{\vX}_{(1)}$ which in this case depicts the relationships of variables that are in the active set of more than one response.  In this case $L_{21}$ and $L_{12}$ correspond to relationships between variables that are active for at least one response, but are not active in all responses.  The sub-matrix $L_{22}$ are the relationships between predictors contained in $\dot{\vX}_{(2)}$, which are the relationships of predictors that are not contained in the active set for more than one response} .\bfblue{(We should discuss the following three sentences at our next meeting. I think this is too general. My understanding is this is about response-predictor interactions. For instance my understanding is $L_{22}$ could contain a predictor that is active for some responses, but not active for others.) \bfred{I tried a little rewrite here}.}   
Finally, define $\vC_{11}=\frac{1}{n}\dot{\vX}_{(1)}^T\dot{\vX}_{(1)}$, $\vC_{22}=\frac{1}{n}\dot{\vX}_{(2)}^T\dot{\vX}_{(2)}$, and $\vC_{21}=\vC^T_{12}=\frac{1}{n}\dot{\vX}_{(2)}^T\dot{\vX}_{(1)}$.

\citet{lassoMsCons} first presented conditions for model selection consistency for lasso. While, \citet{li_10} present model selection results for a lasso-type estimator that includes a penalty for a graph structure in the predictors. The estimator in that paper is similar to the one in this paper, but with $\vA$ fixed and a univariate response. The following conditions are generalizations of conditions presented in those papers. For vectors $\va,\vb \in \Real^d$, let $\mbox{sign}(\va)$ be a vector of the signs of the entries of $\va$, $|\va|$ be the vector formed by taking the absolute values of each entry of $\va$, $\min(\va)$ is the smallest entry of $\va$ and $\va<\vb$ is true if and only if $a_i < b_i$ for all $i \in \{1,\ldots,d\}$.

%, while \citet{li_10} provide model selection consistency results for an estimator similar to the one proposed in this paper, but with $\vA$ fixed and $r=1$. We use similar conditions to derive model selection  
%
%For selection consistency we condition on a selected $\vA$ and as \eqref{vecm_model} is the formulation on \citet{li_10} with $\vA$ known extended to a multivariate regression setting.  We make three additional assumptions which are more restrictive than those needed for high dimensional consistency.  
\begin{condition}
\label{normal_assume} 
Assume $\vepsilon_i \sim N_r(\mathbf{0}_r, \sigma^2\vD_0)$,  are $i.i.d$ and $\vD_0 \in \Real^{r \times r}$ is a diagonal matrix where diagonal values are on the support $(0,1]$ and thus the maximum variance of the errors is $\sigma^2$.  
\end{condition}
%\bfblue{Added dimensions for $\vepsilon_i$ and $D_0$, this is always helpful for me to understand what we are talking about. Please verify that I got them right.}

\begin{condition}
\label{postiveDef}
The minimum eigenvalue of $\vC_{11}$, $k_1$, is positive for all $n$. 
 %Assume $k_1=\rho_{\min}(\vC_{11})>0$. 
\end{condition}

%As defined in \citet{lassoMsCons} the the estimator is strongly sign consistent if 
%$$
%\lim_{n \rightarrow \infty} P[\sign(\dot{\vbeta}(\vA)=\sign(\dot{\vbeta}^*)]=1
%$$
%Following the concepts laid out in \cite{li_10} we define model selection consistency to be equivalent to sign consistency.  We define sign consistency as there exists a $\delta$,$\gamma$ as functions of $n$ that 
%$$
%\lim_{n \rightarrow \infty} P[\sign(\hat{\vbeta}(\vA)=\sign(\vbeta)]=1,
%$$
%where $\hat{\vbeta}(\vA)$ is the solution to \eqref{vecm_model}  and depends on $\gamma$,$\delta$ though the notation does not indicate it.  
%As defined in 

\begin{condition}
\label{mvgc-ic}
%(MVGC-IC):
There exists a fixed $\eta$, that does not change with $n$, such that 

$$
\left|\left(\vC_{21}+\frac{\gamma}{n}\vL_{21}\right)\left(\vC_{11}+\frac{\gamma}{n}\vL_{11}\right)^{-1} \left\{\sign[\dot{\vbeta}^*_{(1)}]+\frac{\gamma}{\delta}\vL_{11}\dot{\vbeta}^*_{(1)}\right\}-\frac{\gamma}{\delta}\vL_{21}\dot{\vbeta}^*_{(1)}\right|\leq \vec{1}-\eta.
$$
\end{condition}

%This now allows us to apply a version of lemma 1 from \citet{li_10_sup}, of which a statement of the lemma and proof can be found in the supplementary material. Using this result we can now provide the statement for selection consistency of the MinPen estimator.   %For completeness we restate that lemma here. \bfblue{I think you moved the Lemma to the supplementary material.}
Condition \ref{mvgc-ic} is multivariate response version of the graph constrained irrepresentable condition from \cite{li_10}, which is an extension of the original irrepresentable condition in \citet{lassoMsCons}. Using these conditions, Lemma 1 from \citet{li_10_sup} and additional conditions on the tuning parameters, we can prove the estimator is model selection consistent. 
\begin{theorem}
\label{modelConsis}
Assume condition \ref{xscale}, \ref{normal_assume}, \ref{postiveDef}, and \ref{mvgc-ic} hold. Define $\bfred{\rho}=min\left[|(\vC_{11}+\gamma/n \vL_{11})^{-1}(\vC_{11}\vbeta_{(1)})|\right]$.% and $W_{\max}$ to be the maximum absolute value of the off diagonal elements of $L$.  
If %$\gamma$ and $\delta$ are choose such that 
$$
\frac{\delta^2}{\log(rp-s)(n+\gamma^2r/k_1)}\rightarrow \infty,
$$
and 
$$
\frac{1}{\rho}\left\{\sqrt{\frac{\log(s)}{nk_1}}+\delta\left\|(\vC_{11}+\gamma/n \vL_{11})^{-1}\sign[\dot{\vbeta}_{(1)}^*]\right\|_{\infty}\right\} \rightarrow 0,
$$
then for any $\vA$, % $\dot{\vbeta}(\vA)$ is a strongly sign consistent estimator for any $\vA$. 
$$
\lim_{n \rightarrow \infty} P\left\{\sign[\dot{\vbeta}(\vA)]=\sign(\dot{\vbeta}^*)\right\}=1.
$$
%
%\begin{itemize}
%\item[(a)]  If $L_{12}=0$,
%\begin{equation*}
%\frac{\delta^2}{n\log(rp-s)} \rightarrow \infty
%\end{equation*}
%or If $L_{12}\neq 0$,
%\begin{equation*}
%\frac{\delta^2}{\log(rp-s)(n+\gamma^2r/k_1)}\rightarrow \infty
%\end{equation*}
%
%\item[(b)] If 
%\begin{equation*}
%\frac{1}{k_1}\left\{\sqrt{\frac{\log(s)}{k_2}}+\delta\|(\vC_{11}+\gamma L_{11})^{-1}\sign(\vbeta_{(1)})\|_{\infty}\right\} \rightarrow 0,
%\end{equation*}
%then the estimator proposed in \eqref{minpen2c} is sign consistent as $n \rightarrow \infty$.  
%
%\end{itemize}
\end{theorem}

Thus the estimator is is strongly sign consistent and therefore model selection consistency holds, see \citet{lassoMsCons} for a more detailed discussion of this relationship. 
%\bfblue{(I would like to discuss the connection between the two assumed rates and our proof. I can sort of make a connection, but it is not immediately clear. I'm also worried that there might be some typos.)}
%\bfblue{Dropped the $L_{12}$ zero or not and only used the stronger condition.}

%\bfblue{If we have probabilities for the above result, like we have for the rate of convergence resutls, then we can get a similar result regarding model selection consistency for $\hat{\vbeta}(\vA)$.}
%\bfred{So there are probabilities to get the bounds which I think are the probabilities of the two conditions.  Could we do that over a union?  Other than that I think we're good on this proof.  }

\subsection{Post Selection Inference}
%Next we present results on inference of our estimators conditioning on the $\hat{\vbeta}$ and $\hat{\vA}$ by solving \eqref{minpen2a}. 
This subsection provides details how to construct confidence intervals for non-zero coefficients of $\hat{\vbeta}(\hat{\vA})$ using the framework provided in \citet{postLee} for lasso estimators. Interpretation of models depends on the variables in the model and we restrict our analysis to inference conditional on the selected model. A consequence of this is $\vbeta^*$ will change depending on the model. Let $Q \subseteq \{1,\ldots,pr\}$ be a subset of selected predictor-response combinations. For simplicity of notation we assume that each response has at least one predictor selected. Define $\tilde{\vX}_Q \in \Real^{nr \times |Q|}$ to be the matrix made from the columns of $\tilde{\vX}$ that belong to $Q$. Define $\vmu = E(\tilde{\vy}) \in \Real^{nr}$, $\tilde{\vX}_Q^+=\left(\tilde{\vX}^T_Q\tilde{\vX}_Q\right)^{-1}\tilde{\vX}_Q^T \in \Real^{|Q| \times nr}$ and
\begin{equation}
\label{psi_truth}
\vbeta^*_Q = \underset{\vbeta \in \Real^{|Q|}}{\mbox{arg min}} E \left|\left| \tilde{\vy}- \tilde{\vX}_Q\vbeta\right|\right|^2_2 = \tilde{\vX}_Q^+\vmu.
\end{equation}
Where $\vbeta^*_Q = (\beta^*_{Q,1},\ldots,\beta^*_{Q,|Q|})^T \in \Real^{|Q|}$, is the true vector of the coefficients conditional on using only the covariates in $Q$. For purposes of post selection inference, we assume that $\tilde{\vy}$ follows a normal distribution of $\tilde{\vy} \sim N(\vmu, \vSigma_{\vepsilon} \otimes \vI_n)$. 

%{\bf Around here or somewhere else we should have a discussion about how large a model we can select. All of Lee's post selection inference work is about inference on coefficients that come from re-fitting a least squares model. I suspect that is what we want to do as well. In that case then we need the number of predictors selected for each response to be smaller than $n$. However, with our ridge-type penalty this may not always be the case. Might be worth doing two things (1) verifying that we can actually select a model where the number of predictors for a given response is larger than $n$ and (2) state that for post-selection inference we assume the user is interested in a sparse model and that the number of predictors selected for each response will be smaller than $n$.}  

Define the vector $\hat{\vs} = (\hat{s}_1,\ldots,\hat{s}_{pr})^T \in \Real^{pr}$, where $\hat{s}_j = \mbox{sign}(\hat{\beta}_j)$ if $\hat{\beta}_j \neq 0$ and $\hat{s}_j \in [-1,1]$ if $\hat{\beta}_j = 0$. As defined in \citet{lassoUnique} for the lasso estimator, for technical convenience we consider the equicorrelation set 
\begin{equation}
\hat{Q} \equiv \{ j \in \{1,\ldots,pr\} \mid |\hat{s}_j| = 1\}.
\end{equation}
The index of nonzero coefficients of $\hat{\vbeta}$ is a subset of $\hat{Q}$ and typically these sets are the same. 

Let $\hat{Q}$ represent the selected model and let $\hat{\vA} \in \Real^{pr(r-1)/2 \times pr}$ be the graph selected from \eqref{minpen2c}. Consider a specific index $j \in Q$, the goal of this work is to define an interval $C_q^{\vA,Q}$ for a given level $\alpha$, such that 
\begin{equation}
P( \beta_{Q,j}^* \in C_j^{\vA,Q} | \hat{Q} = Q, \hat{\vA}=\vA) \geq 1-\alpha.
\end{equation}
See \citet{postLee} for a justification of coverage of post-selection intervals being conditional on the selected model and a review of other approaches to post-selection inference. To derive this interval we consider an arbitrary vector of $\veta_Q \in \Real^{nr}$  examine the conditional distribution of 
\begin{equation}
\veta^T_Q \tilde{\vy} | (\hat{Q}=Q, \hat{\vA}=\vA).
\end{equation}

Note that this definition does include all covariates that have non-zero coefficients, but could potentially include a non-zero coefficient as $\hat{s}_i \in [-1,1]$. %{\bf In Lee's paper they state that this does not happen very often. I'm guessing that result is in \citet{lassoUnique}. Also, we aren't dealing with a lasso estimator, but might be easy to gloss over this. Apparently this ``version'' of $\hat{Q}$ is easier to deal with.} 
For $\hat{\vbeta}$ and $\hat{s}$ to solve \eqref{grace}, conditional on $\hat{\vA}=\vA$, the following sufficient and necessary KKT conditions must be satisfied,
\begin{equation}
\label{kkt1}
\left(\frac{1}{n}\tilde{\vX}^T \tilde{\vX}+ \gamma \vA^T \vA\right) \hat{\vbeta}(\vA) - \frac{1}{n} \tilde{\vX}^T  \tilde{\vy} + \delta \hat{s} = \mathbf{0}_{pr},
\end{equation}
\begin{equation}
\label{kkt2}
\hat{s}_j = \mbox{sign}(\hat{\beta}_j) \mbox{ if } \hat{\beta}_j \neq 0,
\end{equation}
and 
\begin{equation}
\label{kkt3}
\hat{s}_j \in [-1,1] \mbox{ if } \hat{\beta}_j = 0.
\end{equation}
For a given $\vA$, $\hat{\vbeta}(\vA)$ is a unique minimizer of \eqref{grace} and confidence intervals are constructed using these KKT conditions. 

%\subsection{Extension of results from Section 6 of \citet{postLee}}

%\subsubsection{Extension of Theorem 6.1 from \citet{postLee}}

For the following results we need the OLS solution to be well defined and thus need $\tilde{\vX}_Q^T\tilde{\vX}_Q$ to be positive definite.  
\begin{condition}
\label{cond_xq_invert}
The matrix $\tilde{\vX}_Q^T\tilde{\vX}_Q$ is positive definite.
\end{condition}
%\bfblue{Is this a repeat of a condition we have in the model selection consistency part?}\bfred{I don't think so looking at it, as in this case $Q$ is conditional on the model where in the other case (1) is the true active set.}

%First theorem is if we want to condition on the selected model, signs of the non-zero coefficients and graph. 

%We first present a theorem for inference that allows us to condition on the selected model signs of the coefficients, as well as the selected matrix $\hat{\vA}$.  
Define $\ve_j$ as the vector with zeros everywhere, but with a one in the $j$th position. The following theorem presents how to construct confidence intervals conditional on $\hat{Q}$ and $\hat{\vA}$. 

\begin{theorem}
\label{lee_thm61_part0}
Let $F^{[a,b]}_{\mu,\sigma^2}(x)$ be the CDF of a random normal variable with mean $\mu$, variance $\sigma^2$, truncated to the interval $[a,b]$. Let $\veta = \tilde{\vX}(\tilde{\vX}^T \tilde{\vX})^{-1}\ve_j$ and define $L$ and $U$ such that 
\begin{equation*}
F_{L,\veta^T\tilde{\vSigma}\veta}^{[V_{\vs_Q,\vA}^-,V_{\vs_Q,\vA}^+]} (\veta^T\tilde{\vy}) = 1-\frac{\alpha}{2} \mbox{ and } 
F_{U,\veta^T\tilde{\vSigma}\veta}^{[V_{\vs_Q,\vA}^-,V_{\vs_Q,\vA}^+]} (\veta^T\tilde{\vy}) = \frac{\alpha}{2}.
\end{equation*}
Where $V_{\vs_Q,\vA}^-$ and $V_{\vs_Q,\vA}^+$ are constants that depend on $\vs_Q \in \Real^{|Q|}$, $\vA$, $Q$, $\tilde{\vSigma}$, $\tilde{\vX}_Q$ and $\vy$. If Condition \ref{cond_xq_invert} holds then
\begin{equation*}
P\left( \beta_{Q,j}^{*} \in [L,U] | \hat{Q}=Q, \hat{\vs}_{\hat{Q}}=\vs_Q, \hat{\vA}=\vA\right) = 1-\alpha.
\end{equation*}
\end{theorem}
%\begin{proof}
%This follows directly from the proof of Theorem 6.1 in Lee's paper, which is an application of Theorem 9.2.12 from Casella and Berger. 
%\end{proof}
Theorem \ref{lee_thm61_part0} provides a way to construct confidence intervals for a least squares model that accounts for the fact that variable selection was first done by satisfying the KKT conditions of the estimator $\hat{\vbeta}(\hat{\vA})$. The result uses the framework proposed in \citet{postLee}, which provides a more detailed discussion of the issue of post-selection inference and the intuition behind the interval construction. Theorem \ref{lee_thm61_part0} assumes that $\tilde{\vSigma}$ is known, but usually in practice it will need to be estimated. If $p<n$, then the covariance matrix of the residuals from a saturated multivariate regression model could be used.  An alternative approach would be to make some simplifying assumption to make it easier to estimate $\vSigma$,  for instance assume it is diagonal.

\subsection{Algorithm}
\label{normal_alg}

%\bfblue{Maybe save this presentation for the algorithm?}

%The optimization in \eqref{minpen1} is non-convex due to the minimum function.   
To obtain the estimates for the estimator defined in \eqref{minpen1} we propose using the formulation of the estimator that is similar to the optimization defined in \eqref{setup}.  Define the following sets 
\begin{equation*}
\begin{split}
P_l=&\left\{m: \|\vbeta_l-\vbeta_m\|_2^2\leq\|\vbeta_l+\vbeta_m\|_2^2, \|\vbeta_l-\vbeta_m\|_2^2<\|\vbeta_l\|_2^2,  m \in\{1,\ldots,r\} \setminus \{l\}\right\}\\
N_l=&\left\{m: \|\vbeta_l+\vbeta_m\|_2^2<\|\vbeta_l-\vbeta_m\|_2^2, \|\vbeta_l+\vbeta_m\|_2^2<\|\vbeta_l\|_2^2,  m \in \{1,\ldots,r\} \setminus \{l\}\right\}\\
Z_l=&\left\{m: \|\vbeta_l\|_2^2\leq \|\vbeta_l-\vbeta_m\|_2^2,\|\vbeta_l\|_2^2 \leq \|\vbeta_l+\vbeta_m\|_2^2, m \in \{1,\ldots,r\} \setminus \{l\}\right\}\\
\end{split}
\end{equation*}
such that $P_l \cup N_l \cup Z_l =\{1,\ldots,r\} \setminus \{l\}$, $P_l \cap N_l =\emptyset$, $P_l \cap Z_l =\emptyset$, and $Z_l \cap N_l =\emptyset$ $\forall l \in \{1,\ldots,r\}$.  %In addition, assume $m \in P_l$ then $l \in P_m$, and similar statements can be made with regard to $Z_l$ and $N_l$. 
Solving \eqref{minpen1} is equivalent to solving %and \eqref{minpen2a} are equivalent to solving 
\begin{equation}
\label{minpen3}
\begin{split}
\argmin_{\vB \in \Real ^{p \times r}, P_1,N_1,Z_1,\ldots, P_r,N_r,Z_r} &\frac{1}{2n}\sum_{k=1}^r \sum_{i=1}^n(\vy_i-\vx_i^T\vbeta_k)^T(\vy_i-\vx_i^T\vbeta_k) + \delta\sum_{k=1}^r \|\vbeta_j\|_1\\
+& \frac{\gamma}{2}\sum_{k=1}^r\left(\sum_{l \in P_k} \|\vbeta_l-\vbeta_k\|_2^2 + \sum_{m \in N_k} \|\vbeta_m +\vbeta_k\|_2^2 +\sum_{s \in Z_k} \|\vbeta_k\|_2^2\right).
\end{split}
\end{equation}
By using this formulation we are able to propose an iterative algorithm that iterates between estimating  the sets $P_l,  N_l, $ and $Z_l$, for each $l \in \{1,\ldots,r\}$, with the regression coefficients fixed,  and estimating the regression coefficients with the sets fixed.  %This two stage approach can be thought of as a generalization of the algorithms proposed by \cite{witten14},   \cite{price_sherwood},  and \cite{price_molstad_sherwood} who propose iterating between estimating parameters of interest and the using the well studied k-means algorithm to define sets.   
%If   $P_l,  N_l, $ and $Z_l$, $l=1,\ldots,r$  are known the optimization in \eqref{minpen3} reduces to a convex optimization problem that can be solved using coordinate descent,  specifically solutions to the well studied elastic net and graph constrained regression problems.  
To initialize the algorithm estimates of $\vB$ or  $P_l,  N_l, $ and $Z_l$, $l=1,\ldots,r$ are needed. We define initial values for the coefficients of the $c$th response are set by  

\begin{equation}
\label{initEst}
\hat{\vbeta}^{(0)}_c=\argmin_{\vbeta_c\in \Real^p} \frac{1}{2n} \sum_{i=1}^n (y_{ic}-\vx_i^T\vbeta_c)^2+\delta\|\vbeta_c\|_1 +\gamma\|\vbeta_c\|_2^2, 
\end{equation}
and define $\hat{\vB}^{(w)}=\left(\hat{\vbeta}_1^{(w)},\ldots,\hat{\vbeta}^{(w)}_{r}\right)$ to be the $w$th iterative estimate of $\vB^*$.  Note we will also use $\hat{\vbeta}^{(w)}$ to represent the vectorized version of $\hat{\vB}^{(w)}$, consistent with previous notation.   Given a fixed $(\delta, \gamma)$ we propose the following algorithm as a solution to \eqref{minpen3}.
\begin{enumerate}
\item Initialize $\hat{\vbeta}_1^{(0)},\ldots, \hat{\vbeta}_r^{(0)}$ as defined in \eqref{initEst}. % and $\gamma, \delta>0$.
\item   For the $w$th iteration, where $w>0$, repeat the following steps until the estimated sets $P_l,  N_l$ and $ Z_l$ do not change from iterate $w-1$ to iterate $w$ for all $l \in \{1,\ldots,r\}$.
\begin{enumerate}
\item Holding $\hat{\vB}^{(w-1)}$ fixed obtain estimates for $\hat{P}^{(w)}_k,\hat{N}^{(w)}_k,\hat{Z}^{(w)}_k$ for all $k \in \{1,\ldots,r\}$ by solving the optimization
\begin{equation}
\label{set_update}
\begin{split}
\underset{P_k,N_k,Z_k \forall k \in \{1,\ldots,r\}}{\mbox{minimize}}& \sum_{k=1}^r\left(\sum_{l \in P_k} \|\hat{\vbeta}^{(w-1)}_l-\hat{\vbeta}^{(w-1)}_k\|_2^2 + \sum_{m \in N_k} \|\hat{\vbeta}^{(w-1)}_m +\hat{\vbeta}^{(w-1)}_k\|_2^2 +\sum_{s \in Z_k} \|\hat{\vbeta}^{(w-1)}_k\|_2^2 \right). 
\end{split}
\end{equation}

\item  Holding $\hat{P}^{(w)}_k,\hat{N}^{(w)}_k,\hat{Z}^{(w)}_k$ for all $k \in \{1,\ldots,r\}$ fixed obtain the estimate $\hat{\vB}^{(w)}$ by solving the optimization
\begin{equation}
\label{minpen3_alg}
\begin{split}
\argmin_{\vB \in \Real ^{p \times r}} &\frac{1}{2n}\sum_{k=1}^r \sum_{i=1}^n(\vy_i-\vx_i^T\vbeta_k)^T(\vy_i-\vx_i^T\vbeta_k) + \delta\sum_{k=1}^r \|\vbeta_j\|_1\\
+& \frac{\gamma}{2}\sum_{k=1}^r\left(\sum_{l \in \hat{P}^{(w)}_k} \|\vbeta_l-\vbeta_k\|_2^2 + \sum_{m \in \hat{N}^{(w)}_k} \|\vbeta_m +\vbeta_k\|_2^2 +\sum_{s \in \hat{Z}^{(w)}_k} \|\vbeta_k\|_2^2\right).
\end{split}
\end{equation}
\end{enumerate}
\end{enumerate}

The update steps, shown in \eqref{set_update} and \eqref{minpen3_alg} respectively,  break the non-convex problem into two problems that can be solved directly with well studied solutions.  The optimization in \eqref{set_update} can actually be solved directly in a single pass,  where the set assignments to the $w+1$ iteration  for $l=1,\ldots, r$  are defined as
\begin{equation*}
\begin{split}
\hat{P}^{(w+1)}_l=&\left\{m: \|\hat{\vbeta}^{(w)}_l-\hat{\vbeta}^{(w)}_m\|_2^2\leq\|\hat{\vbeta}^{(w)}_l+\hat{\vbeta}^{(w)}_m\|_2^2, \|\hat{\vbeta}^{(w)}_l-\hat{\vbeta}^{(w)}_m\|_2^2<\|\hat{\vbeta}^{(w)}_l\|_2^2,  m \in\{1,\ldots,r\} \setminus \{l\}\right\},\\
\hat{N}^{(w+1)}_l=&\left\{m: \|\hat{\vbeta}^{(w)}_l+\hat{\vbeta}^{(w)}_m\|_2^2<\|\hat{\vbeta}^{(w)}_l-\hat{\vbeta}^{(w)}_m\|_2^2, \|\hat{\vbeta}^{(w)}_l+\hat{\vbeta}^{(w)}_m\|_2^2<\|\hat{\vbeta}^{(w)}_l\|_2^2, m \in \{1,\ldots,r\} \setminus \{l\}\right\},\\
\hat{Z}^{(w+1)}_l=&\left\{m: \|\hat{\vbeta}^{(w)}_l\|_2^2\leq \|\hat{\vbeta}^{(w)}_l-\hat{\vbeta}^{(w)}_m\|_2^2,\|\hat{\vbeta}^{(w)}_l\|_2^2 \leq \|\hat{\vbeta}^{(w)}_l+\hat{\vbeta}^{(w)}_m\|_2^2, m \in \{1,\ldots,r\} \setminus \{l\}\right\}.\\
\end{split}
\end{equation*}
Note that this update requires calculating $r^2$ terms and performing $r(r-1)$ comparisons, but the sets are deterministic and no iterates are needed.   

The optimization in \eqref{minpen3_alg} can be solved using a gradient descent algorithm where each $\beta_{mk}$ is solved for iteratively with all other regression coefficients held fixed.  Let $S(x,a) = \mbox{sign}(x)(|x|-a)_+$ be the soft-thresholding operator. To solve \eqref{minpen3_alg} we use a coordinate descent algorithm where the update for the $m$th predictor for the $k$th response, $\bar{\beta}_{mk}$, is

$$
\bar{\beta}_{mk}=\frac{S\left(n^{-1}\{\sum_{i=1}^n x_{im}y_{ik}-\sum_{j=1, j\neq m}^p (\sum_{i=1}^n x_{im}x_{ij})\bar{\beta}_{jk} +\gamma H_{mk}\},\delta/2\right)}{n^{-1}\left(\sum_{i=1}^nx_{im}^2+\gamma\left\{(r-1)+\sum_{g=1}^r I\left[k \in \hat{P}_g^{(w)}\right] +\sum_{v=1}^r I\left[k \in \hat{N}_v^{(w)}\right] \right\}\right)},
$$
%\bfblue{(Problem in the above numerator there is a $\{$, but no $\}$.)}
where
$$
H_{mk}=\sum_{l \in \hat{P}_k^{(w)}} \bar{\beta}_{ml} -\sum_{h \in \hat{N}_k^{(w)}} \bar{\beta}_{mh} +\sum_{g=1}^r \bar{\beta}_{mg}I\left[k \in \hat{P}_g^{(w)}\right] -\sum_{v=1}^r \bar{\beta}_{mv}I\left[k \in \hat{N}_v^{(w)}\right],
$$
and $\bar{\beta}_{jv}$ are the current iterates of the coordinate descent algorithm.   The algorithm iterates through all $m \in \{1,\ldots,p\}$ and $k \in \{1,\ldots,p\}$ in a similar fashion to other coordinate descent algorithms and converges with the change in the estimates across iterations is small. We propose selecting $\delta$ and $\gamma$ using $k$-fold cross validation minimizing the validation residual sum of squares. The proposed coordinate descent update is similar to GRACE, but differs due to structured use of the minimum penalty across multivariate responses. % different due to the minimum penalty and multivariate response.%ACE estimators proposed by \cite{li_08} and \cite{li_10}, but different due to the minimum penalty and multivariate response.} %We also note that our update is related to a multivariate version of the GRACE estimators proposed by \cite{li_08} and \cite{li_10} but with different properties as the minimum penalty we propose is different from the graph Laplacian used in previous works.   %Though in implementation, the solution to be \eqref{minpen3_alg} can still be solved using standard software such as the glmnet algorithm proposed by \cite{friedman2008} and is widely available in open source software such as R and Python.     

While this algorithm is related to the iterative algorithms proposed by \cite{witten14}, \cite{price_sherwood},  and \cite{price_molstad_sherwood}, there are some differences.  First, the penalties investigated by those authors are considered \emph{cluster fusion penalties}. Second, the algorithms proposed in those papers are two stage procedure that require a solution to the well studied $k$-means problem, which can be unreliable and computationally burdensome in high-dimensional settings. The proposed algorithm provides an efficient algorithm that is guaranteed to find the optimal sets of the responses given the coefficients. %Our penalty does not require an iterative algorithm to define sets.  The second, is a related point,  which is that our algorithm shows that these iterative procedures generalize outside of the \emph{cluster fusion penalized} framework, which to this point has only been used in the context of $k$ means clustering, thus providing a standard approach for methods that require sets to be defined while estimating regression parameters.  
An alternative approach would be an exhaustive search of all possible $\vA \in \As$ which would provide a global minimizer.  However, this becomes computational intractable for medium or large $r$ as there are $3^{r(r-1)}$ combinations that must be searched before tuning parameter selection is considered. % (e.g.  $r=3$ has 729 cases to consider).   
The iterative method we propose is not constrained in this way but is not guaranteed to achieve a global solution to the non-convex optimization problem.   %Finally, we believe our algorithm provides a framework for generalization beyond our choice of a minimum penalty.  Inside of the minimum function we have use ridge type penalties, but it may be reasonable to study fused lasso penalties, or develop other penalties of interest that could be related to the truncated lasso \citep{shen2012likelihood,tibs_14}.
The proposed algorithm can be easily adjusted to accommodate if other penalty functions, besides the proposed ridge fusion, are used within the minimum function.

\section{Extension to  Multiple Binomial Responses}
\label{bin_res}
%\section*{Extending to the GLM Setting}
\subsection{Method}

This section details how to extend the proposed method to the setting where conditional on the predictors each response follows a binomial distribution. The focus will be on the binomial setting, but the method presented in this section can be generalized to multiple responses from the same exponential family with different mean functions.
%We next extend this where each response follows the same exponential family,  with different mean functions.   For now let's restrict our discussion to just the Binomial logistic setting as there are various applications of interest across many fields including precision medicine, engineering, and marketing with the understanding the this extension could be taken more generally to any exponential family setting.  
Define $\vtheta_k=(\alpha_k,\vomega_k^T)$, where $\alpha_k$ is a response specific intercept, and $\vu_i=(1,\vx_i)$ be the $p+1$ dimensional vector of covariates for the $i$th observation.  Let $y_{ik}$ be a realization of the random variable 

$$
Y_{ik}\sim \Bin(n_{ik},\pi_{ik}), \mbox{ where } 
\pi_{ik}=\frac{\exp(\vu_i^T\vtheta_k)}{1+\exp(\vu_i^T\vtheta_k)}.
$$
%for $i=1,\ldots, n$ and $k=1,\ldots, r.$, where $x_i$ is a $p$-dimensional set of covariates and $\beta_k$ is a set of regression coefficients for the $k$th response.  

We define the penalized likelihood for the minimum penalized model as

\begin{equation}
\label{BinPen}
\begin{split}
\sum_{k=1}^r\sum_{i=1}^n& y_{ik}\vu_i^T\vtheta_k-n_{ik}\log\left[1+\exp(\vu_i^T\vtheta_k)\right]+\delta\sum_{k=1}^r\|\vomega_k\|_1\\
+&\frac{\gamma}{2}\sum_{l=1}^{r}\sum_{m=1,l\neq m}^r\min\left(\| \vtheta_l-\vtheta_m\|_2^2, \|\vtheta_l+\vtheta_m\|_2^2, \| \vtheta_l\|_2^2\right).
\end{split}
\end{equation}
%where the notation $\vtheta_k^{(-1)}$ represents the vector $\vtheta_k$ with the first element removed.  
%Note a change from the Gaussian setting as we penalize differences that include the intercept term similar to that of \cite{price_sherwood}.  
The lasso penalty does not include the intercept, because we assume that an intercept is part of the model for each response. However, the intercept is part of the penalty in the minimum function because we wish to group responses based on the relationships between fitted values.

\subsection{Algorithm}
We propose solving \eqref{BinPen} by approximating it with a penalized quadratic function similar to the glmnet algorithm \citep{glmnet}.  Define, 
\begin{equation}
g(\pi_{ik})=\log\left(\frac{\pi_{ik}}{1-\pi_{ik}}\right)=\vu_i^T\vtheta_k.%\tilde{\vx}_i'\tilde{\vbeta}_k.
\end{equation}

To implement our penalized quadratic approximation we define the following,
\begin{eqnarray}
z_{ik}&=&g(y_{ik})=g(\pi_{ik})+\frac{y_{ik}-\pi_{ik}}{\pi_{ik}(1-\pi_{ik})},\\
w_{ik}&=&\pi_{ik}(1-\pi_{ik}),\\
\label{quad_approx}
-l_{k}(\vtheta_k)&=&\sum_{i=1}^nw_{ik}(z_{ik}-\vu_i^T\vtheta_k)^2.%\sum_{i=1}^nw_{ik}(z_{ik}-\vx_i'\vbeta_k)^2.
\end{eqnarray}
Note that $z_{ik}$ is the first order Taylor approximation of $g(y_{ik})$, and that $w_{ik}$ is the conditional 
variance of $z_{ik}$ given $\vu_i$. Define $\vZ_k = (z_{1k},\ldots,z_{nk})^T \in \Real^n$ and $\vW = (w_{1k},\ldots,w_{nk})^T \in \Real^n$. 

%Then we define the minimum penalty binomial (BinMinPen) model as \bfblue{(A little confused by what is being said here. My understanding, is that this is an approximation of \eqref{BinPen}. From my reading we should be defining the minimizer of \eqref{BinPen} as BinMinPen.)} \bfred{Correct}

%\bfred{To obtain the estimates for the estimator defined in \eqref{BinPen} we propose solving the optimization}
%\bfblue{The estimator 
%\begin{equation}
%\label{BinApprox}
%\begin{split}
 %&\argmin_{\vtheta_k \in \Real^{p+1},  P_k,N_k,Z_k \forall k \in \{1,\ldots,r\} }\sum_{k=1}^r -l_{k}(\vtheta_k) +\delta\sum_{k=1}^r\|\vtheta^{(-1)}_k\|_1\\&
 %+\frac{\gamma}{2}\sum_{l=1}^{r}\left( \sum_{m \in P_l} \|\vtheta_m-\vtheta_l\|_2^2, \sum_{h \in N_l } \|\vtheta_m+\vtheta_l\|_2^2,  \sum_{v \in Z_l} \|\vtheta_l\|_2^2 \right),
 %\end{split}
%\end{equation}
%where $P_l, N_l$ and $Z_l$ $l=1,\ldots,r$ are defined the same as in Section \ref{norm_method} but use $\vtheta_1,\ldots,\vtheta_k$ rather than $\vbeta_1,\ldots\vbeta_k$.  
Let $P_l, N_l$ and $Z_l$ be defined as in Section \ref{norm_method}, but now with respect to $\vtheta_l$ instead of $\vbeta_l$. We propose an iterative algorithm similar to that proposed in Section \ref{normal_alg} to minimize \eqref{BinPen}. %to solve the optimization presented in \eqref{BinApprox}.  

Initial estimates of  $\vtheta_l$ or $P_l,  N_l, $ and $Z_l$ are needed for all $l \in \{1,\ldots,r\}$. Similar to the method proposed in Section \ref{normal_alg} we propose initializing the regression coefficients for each of the $r$ responses separately using the elastic net estimator, which we define as $\hat{\vtheta}_k^{(0)}$. %\bfblue{(We imply that an initial estimate is required for the sets, but are not providing any details about the initial estimates.)} \bfred{doesn't the sentence before what you wrote do that?}% for all $k=1,\ldots,  r$.  
Given a fixed $(\delta, \gamma)$ we propose the following algorithm as a solution to \eqref{quad_approx}.
\begin{enumerate}
\item Initialize $\hat{\vtheta}_1^{(0)},\ldots, \hat{\vtheta}_r^{(0)}$.% and $\gamma, \delta>0$.
\item   For the $w$th iteration where $w>0$, repeat steps until the estimated sets $P_l$, $N_l$ and $Z_l$ do not change from iterate $w-1$ to iterate $w$:
\begin{enumerate}
\item Holding $\hat{\vtheta}_k^{(w-1)}$ fixed for all $k \in \{1,\ldots,r\}$, obtain $\hat{P}^{(w)}_l$, $\hat{N}^{(w)}_l$ and $\hat{Z}^{(w)}_l$ for all $l \in \{1,\ldots,r\}$ by solving the optimization
\begin{equation}
\label{set_updateBin}
\begin{split}
\underset{P_1,N_1,Z_1,...,P_r,N_r,Z_r}{\mbox{minimize}}& \sum_{k=1}^r\left(\sum_{l \in P_k} \|\hat{\vtheta}^{(w-1)}_l-\hat{\vtheta}^{(w-1)}_k\|_2^2 + \sum_{m \in N_k} \|\hat{\vtheta}^{(w-1)}_m +\hat{\vtheta}^{(w-1)}_k\|_2^2 +\sum_{s \in Z_k} \|\hat{\vtheta}^{(w-1)}_k\|_2^2 \right). 
\end{split}
\end{equation}

\item  Holding $\hat{P}^{(w)}_k,\hat{N}^{(w)}_k,\hat{Z}^{(w)}_k$ fixed for all $k \in \{1,\ldots, r\}$ obtain the estimates of $\hat{\vtheta}_k^{(w)}$ for all $k \in \{1,\ldots,r\}$ by solving the optimization
\begin{equation}
\label{BinUpdate_alg}
\begin{split}
 &\argmin_{\vtheta_k \in \Real^{r}, \forall k \in \{1,\ldots,r\}}\sum_{k=1}^r -l_{k}(\vtheta_k) +\delta\sum_{k=1}^r\|\omega_k\|_1\\&
 +\frac{\gamma}{2}\sum_{l=1}^{r}\left( \sum_{m \in \hat{P}^{(w)}_l} \|\vtheta_m-\vtheta_l\|_2^2, \sum_{h \in \hat{N}^{(w)}_l } \|\vtheta_h+\vtheta_l\|_2^2,  \sum_{v \in Z^{(w)}_l} \|\vtheta_l\|_2^2 \right).
 \end{split}
\end{equation}
\end{enumerate}
\end{enumerate}

Similar to the algorithm in Section \ref{normal_alg}, this could be replaced with an exhaustive search algorithm for the global minimum. However, even for small $r$ this will be computationally burdensome. The step for estimating the sets is identical to the previous algorithm and thus once again if the coefficients are considered given there is an easily derived global solution for the sets, unlike the K-means algorithm which has been used in similar problems.

For the update  in \eqref{BinUpdate_alg}  we propose using a proximal gradient descent method similar to the glmnet algorithm proposed by \cite{friedman2008}.  Let $\hat{w}_{ik}$ and $\hat{z}_{ik}$ be the most recent estimates of $w_{ik}$ and $z_{ik}$. The update of the $j$th variable of the $m$th response of the coefficients for the proposed proximal gradient descent method is
$$
\bar{\theta}_{jm}=\frac{S\left(\sum_{i=1}^n \hat{w}_{ik}\{x_{ij}\hat{z}_{ik}-\sum_{l=1, l\neq j}^p u_{il}u_{ij}\bar{\theta}_{lm}\} +\gamma M_{jm},I(j\neq 1)\delta/2\right)}{\left(\sum_{i=1}^nu_{ij}^2+\gamma\{(r-1)+\sum_{g=1}^r I(m \in \hat{P}_g^{(w)}) +\sum_{v=1}^r I(m \in \hat{N}_v^{(w)}) \}\right)},
$$
where
$$
M_{jm}=\sum_{l \in \hat{P}_m^{(w)}} \bar{\theta}_{jl} -\sum_{h \in \hat{N}_m^{(w)}} \bar{\theta}_{jh} +\sum_{g=1}^r \bar{\theta}_{jg}I(m \in \hat{P}_g^{(w)}) -\sum_{v=1}^r \bar{\theta}_{jv}I(m \in \hat{N}_v^{(w)}).
$$
The algorithm iterates through all $j \in \{1,\ldots,p\}$ and  $m \in \{1,\ldots,r\}$ until convergence  and then updates the quadratic approximation and continues to solve the IRWLS optimization until convergence.  This entire process is the solution to \eqref{BinUpdate_alg}.  To select $\delta$ and $\gamma$ we propose using $k$-fold cross validation minimizing validation likelihood loss.%, similar to the method proposed by \cite{price_sherwood}.

\section{OLS Simulations}
\label{olssim}
\subsection{Data Generating Model and Evaluation Metrics}
\label{data_gen_ols}
In this section we investigate the performance of the MinPen estimator defined by \eqref{minpen1}, the MinPen estimator with the correct indices matrix $\vA$ known \emph{a priori} (T-MinPen), the multivariate cluster elastic net estimator (MCEN) \cite{price_sherwood},  the separate elastic net estimator (SEN) which fits the elastic net on each response using common tuning parameters, and 
%\begin{equation}
%\hat{\vB}_{SEN}=\argmin_{\vB \in \Real^{r\times p}} \frac{1}{2n}\sum_{k=1}^r \sum_{i=1}^n (y_{ik}-\vx_i^T\vbeta_k)^2+\delta\sum_{m=1}^r \|\beta_m\|_1 +\gamma \sum_{v=1}^r\sum_{j=1}^p \beta_{vj}^2,
%\end{equation}
joint elastic net estimator (JEN) which utilizes a group penalty on the same predictor variable across all responses \citep{friedman2008}. 

The SEN and JEN estimators are fit using the \texttt{glmnet} package in R.   Tuning parameters are selected using a test train procedures, where each model is trained on $n=100$ observations and then evaluated on a different set of $100$ test observations.  The selected tuning parameters minimize the predicted sum of squares error on the test observations.  We establish the T-MinPen estimator as a baseline for the MinPen estimator when the relationships would be known by 
a practitioner prior to fitting the model.  %We note that these relationships are not exact and are inferred as the sparsity included in the model will have implications the strength of these relationships. \bfblue{(I don't understand the previous sentence, but I suspect is is related to my question of how do we define the true structure for T-MinPen.)}

Let  $\tilde{\vSigma}_X \in \Real^{4 \times 4}$ with entries $\tilde{\sigma}_{ii} = 1$ and $\tilde{\sigma}_{ij} = \rho$, for $i\neq j$. The covariates are generated by $\vx_i \sim N(\mathbf{0}_p,\vSigma_x)$, where $\vSigma_x$ is a block diagonal matrix with $p/4$ blocks of $\tilde{\vSigma}_X$ with $\rho=0.7$ on the diagonal blocks and all other entries are set to 0.

Define that data generating model to the be same as proposed in \eqref{lin_mod}
such that $\vepsilon_i \sim N_{15}(\mathbf{0}_{15},\vI_{15})$.  In all simulations we perform 100 replications where each replication consists of using training data set with $n=100$ and tuning parameters are selected using predicted squared error loss on 100 independently sampled observations.  To evaluate the model we generate an independent validation  set consisting of 1000 observations and calculate the average squared prediction error (SPE) %which we define as \bfblue{(Below is not an average, but we are defining it as one. I also think we are talking about some training/testing split here, but some more details are needed)}
$$
\frac{1}{15000}\sum_{k=1}^{15}\sum_{i=1}^{1000} (y_{ik}-\hat{y}_{ik})^2,
$$
where $y_{ik}$ is the $k$th element of $\vy_i$ and $\hat{y}_{ik}$ is a prediction for the $i$th observation and $k$th response.  We also use the mean square error (MSE) of the estimators of the regression coefficient matrix $\vB$, which we defined as 

$$
\frac{1}{15p}\sum_{k=1}^{15} \| \hat{\vbeta}_k-\vbeta_k^*\|_2^2.
$$
We also report on the number of true variables and false variables selected through true positive (TP) and false positive (FP) rates respectively over the 100 replications.% \bfblue{(Earlier we say 50, here we say 100. Or do the two represent different things?)}

\subsection{Block Structure in Regression Coefficients}
\label{block_ols}
In this simulation we will investigate how the methods perform when encountering a structured set of regression coefficients where 

$$
\vB^*=\left(\begin{array}{lll}
\vDelta_{10}(\eta,\lambda) & \vec{0}_{10} & \vec{0}_{10} \\ 
\vec{0}_{10} & \vDelta_{10}(\eta,\lambda) & \vec{0}_{10} \\ 
\vec{0}_{10} & \vec{0}_{10} & \vDelta_{10}(\eta,\lambda) \\ 
\vec{0}_{p-30} & \vec{0}_{p-30} & \vec{0}_{p-30}
\end{array} \right).
$$
We define the regression coefficients for subsets of responses as $\vDelta_q(\eta,\lambda)=(-\veta_q-\lambda, \veta_q, \veta_q+\lambda,-\veta_q-2\lambda,\veta_q+3\lambda)\in \Real^{q \times 5}$ where $\veta_q$ is a $q$-dimensional vector with each element set equal to $\eta$ and $\lambda$ is a constant.  

%This setting will highlight the ability of each method to detect the appropriate relationships between responses as well as identify variables used on different subsets of predictors. 
The proposed method should identify both the positive and negative relationships across the responses and leverage that information to improve prediction accuracy. Of the methods we compare to, only MCEN can identify relationships and it is limited to only identifying positive relationships. We investigate the settings where $(\eta,\lambda) \in \{0.5,1.0\} \times \{0.02, 0.05,0.10\}$ for $p=40, 100$, and $300$.  %A similar setting was explored by \cite{price_sherwood} except only positive relationships were used. 
Figure \ref{fig:ols1} presents the results of the simulation for $p=300$, while the results for $p=40$ and $100$ are shown in the supplemental material.  The results show that MCEN and MinPen perform competitively in MSE and PSE for all values of $\eta$, while JEN is superior when $\gamma=1$.  With respect to true and false positive rates, MinPen out performs or performs as well as the other approaches regardless of the values of $\eta$.  It may be somewhat surprising that MCEN would perform well in this setting, but after investigation MCEN selects two clusters, a positive and negative cluster.  As the effect size is consistent across each of these models, MCEN is competitive with MinPen though does tend to have a higher false positive rate. % even though different variables are found to be important it is intuitive that MCEN will perform well, though MinPen out performs all methods in model identification.  

\begin{figure}
\begin{center}
\includegraphics[height=3in, width=5in]{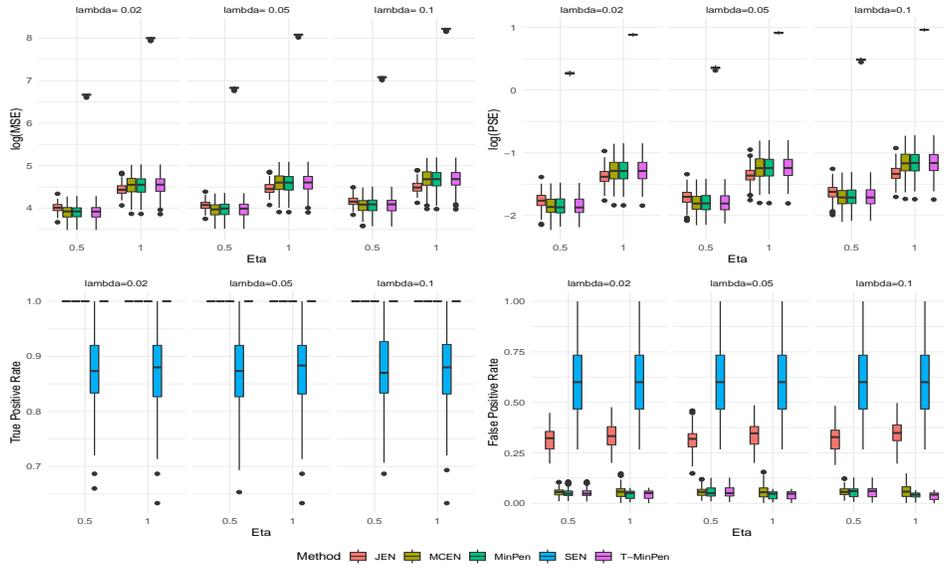}
\end{center}
\caption{Results of the simulation described in Section \ref{block_ols} for the case of $p=300$. }
\label{fig:ols1}
\end{figure}
The results presented show with respect to SPE all methods except for SEN perform similarly for all values of $p$, $\eta$, and $\lambda$.  A similar results is found for both MSE and TP, with MinPen having a slight advantage for smaller $p$.  The real differentiation between methods is shown when comparing the FP rate where we see MCEN and MinPen perform as well as T-MinPen and out perform all competitors  SEN and JEN in the case of $\eta=0.5$.  In the case of $\eta=1.0$ MinPen is able to perform closely to T-MinPen.  For larger values of $\lambda$ we see a much more consistent result in FP in MinPen than MCEN.  %An interesting result in this case is the ability of MCEN to perform competitively with MinPen,  this comes from the simulation design which allows MCEN to detect two clusters related to shared information in predictions, which is why we see degrading performance in the FP rate rather than other metrics. %\bfblue{(Why no metric for set comparison? I would think MinPen would work better than MCEN in that setting.)}\bfred{because there aren't distinct sets to compare,  we like to think things are positively and negative relationships but there aren't distinct groups like when there was MCEN, it's estimation of the matrix $A$ and it got to complex to discuss the measures on that. }

\subsection{ Overlapping Variables in Regression Coefficients}
\label{overlap_ols}

We next investigate a simulation where there is overlap between the non-zero variables in each coefficient vector.  We define $\vB^*$ to be be structured such that the regression coefficients for the $k$th response, with $r=15$, are defined as  
$$
\vbeta^*_k=\left(   \begin{array}{l}
\vec{0}_{v(k-1)} \\ 
(-1)^k\vec{0.5}_{10} \\ 
\vec{0}_{p-v(k-1)-10}
\end{array}  \right),
$$

We investigate the settings $(v,p)\in \{0,2, 4\} \times \{100,300\}$ over 100 replications. Figure \ref{fig:ols2}  presents the case of $p=300$, while the results for $p=100$ are available in the supplemental material.   When $v=0$ each predictor is either active or inactive for all 15 responses. In this setting JEN out performs competitors in all metrics for every $p$ studied, because it uses a group lasso penalty for each predictor across the fifteen different responses. % which is unsurprising given the structure promoted by JEN.  
As the amount of overlap increases MCEN and MinPen methods perform the best with regard to MSE and PSE.  Similar to the simulations presented in Section \ref{block_ols} T-MinPen and MinPen out preform competitors finding fewer false positive variables.  Thus using MinPen is superior as the method is able to detect the underlying structure in the regression coefficients more often than competitors.  %Again, MCEN is able to leverage the fact that both positive and negative responses exist, and is able to detect the two response groups to improve prediction accuracy. %but unable to detect important variables as they are not shared between responses. \bfblue{(Do we provide any evidence for what follows but in the previous sentence?)}

\begin{figure}
\begin{center}
\includegraphics[height=3in, width=5in]{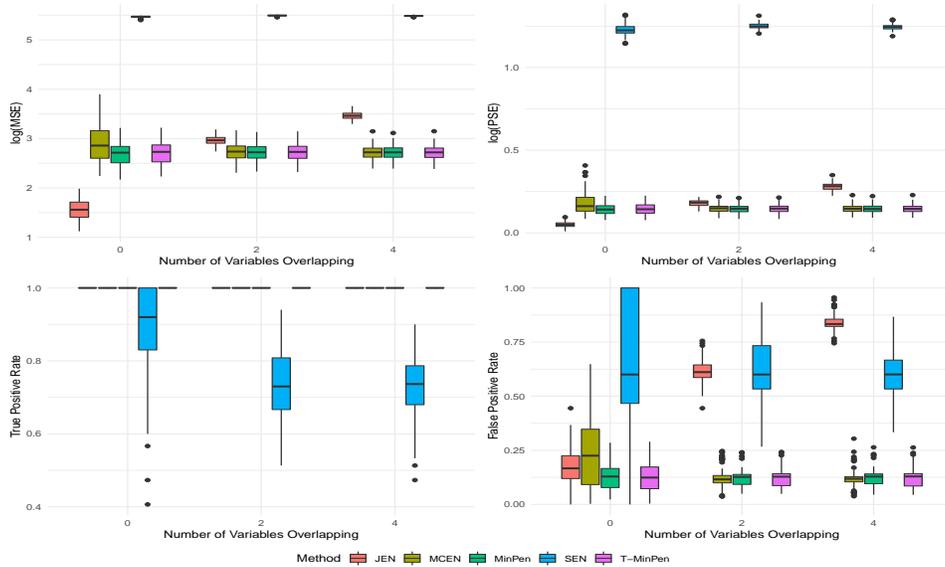}
\end{center}
\caption{Results of the simulation described in Section \ref{overlap_ols}  for the case of $p=300$}
\label{fig:ols2}
\end{figure}

\section{Binomial Simulations}
\label{BinSims}

Next we study the impact of the minimum penalty methodology in binomial logistic regression 
setting and compare it to SEN and MCEN which also have binomial logistic regression cases for the 
respective methods.  We will also compare against T-MinPen which is a version of MinPen where we fix the true $\vA$ to be known by a practitioner \emph{a priori}.   We generate $\{\vx_i\}_{i=1}^n$ in the exact same way as proposed in Section \ref{data_gen_ols}.  The responses are generated by
$$
y_{ik}\sim Bin(1,\pi^*_{ik}) \mbox{ where } 
\pi^*_{ik}=\frac{\exp(\vx_i^T\vbeta^*_k)}{1+\exp(\vx_i^T\vbeta^*_k)}.
$$
In this simulation study we define $\vbeta^*_k$ $k \in \{1,\ldots,r\}$ to be equivalent to the  $k$th column 
of the $\vB^*$ studied in section \ref{block_ols} for $p=40, 100,$ and $300$ respectively. Again we 
study $(\eta,\lambda) \in \{0.5,1.0\} \times \{0.02, 0.05,0.10\}$.  For each of the 100 iterations 
tuning parameters are selected using a train test procedure with 100 observations in the training set, and 100 observations in the test set.   A validation of 1000 observations is to used 
evaluate methods using Kullback-Leibler divergence (KL), that is % which we define as

$$
\sum_{k=1}^{15}\sum_{i=1}^{1000} \left\{\log\left(\frac{\hat{\pi}_{ik}}{\pi^*_{ik}}\right)\hat{\pi}_{ik}+\log\left(\frac{1-\hat{\pi}_{ik}}{1-\pi^*_{ik}}\right)(1-\hat{\pi}_{ik})\right\},
$$ 

where $\hat{\pi}_{ik}$ is the resulting estimated probability.  We also compare methods using  MSE, TP, and FP.  

Figure \ref{Bin300} presents the results for the case of $p=300$.  The results for $p=40$ and $100$ are available in the supplementary material.  The results show that MinPen performs as well as TMinPen and out performs the other competitors with respect to MSE and KL.  We see that  with regard to variable selection, MinPen is comparable, if not better, than the other methods with respect to TP and outperforms all approaches, even TMinPen, with respect to FP.  TMinPen performs better the MinPen on TP but worse on FP indicating it is selecting more variables than necessary.  %This is an important result as it shows that even if MinPen is isn't selecting variables as well as when the truth is known, it does not impact KL in this setting.  

\begin{figure}
\begin{center}
\includegraphics[height=3in, width=5in]{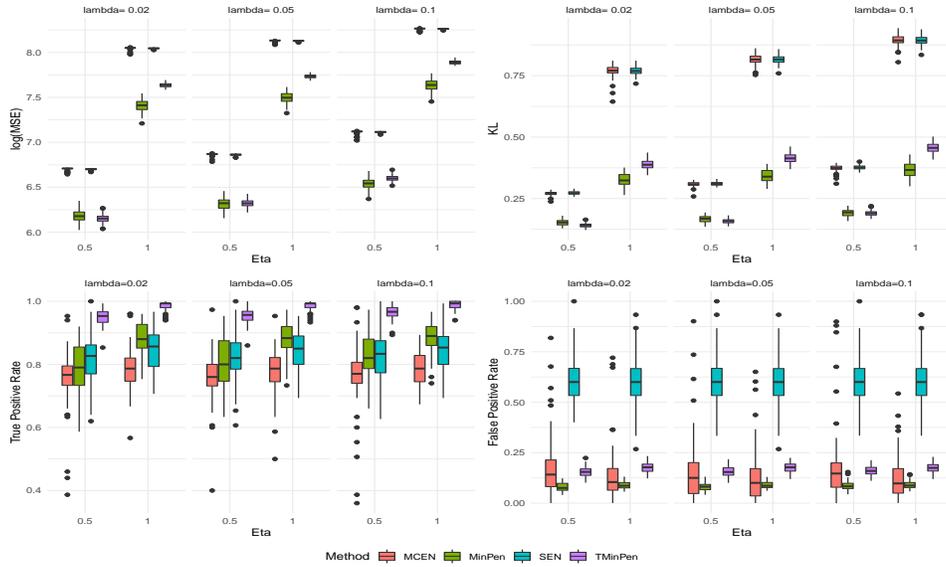}
\end{center}
\caption{Results of the simulation described in Section \ref{BinSims} for $p=300$.}
\label{Bin300}
\end{figure}

\section{Applied Examples}
\label{applications}

\subsection{Genomics Data}
In this section we compare the post model selection inference approach outlined in this paper, after using the proposed least squares method, with the post model selection inference approach proposed by \citet{postLee}, after using lasso for model selection. The models are fit to data analyzed by \citet{genomicData}, who collected demographic, birth and gene expression data from 72 postpartum women and their newborns, but our analysis is limited to 64 of the women, 65 had complete data and one was dropped due to outliers in the demographic data. Four response variables are modeled: placental weight, newborn weight and two measures of cotinine level, one from the mother's peripheral blood and a second from the umbilical cord. The predictors are Smoking status, mother's age, mother's BMI, parity, gestational age and 33 gene expression probes. The 33 probes were selected by taking the absolute value of the correlation for each response and the 24,526 probes measured in the study and using a union of the top ten for each response. 

The lasso models are fit separately for each response. For both methods tuning parameters are selected to minimize the mean squared prediction error from five folds cross validation. For post selection inference, $\tilde{\vSigma}$ is estimated using the covariance matrix of residuals from the full multivariate linear regression model. The four responses are related. First, the two cotinine measurements are measurements of the same variable at nearly the same time, but from different samples. In addition, cotinine levels are high in people who smoke and it has been shown that lower cotinine levels are associated with larger birth weights \citep{cotBirthWt}. In addition, \citet{WANG2014437} found both smoking has a negative relationship with both birth and placenta weights. Finally, newborn and placental weights are positively correlated \citep{linearPlaBrth}. The analysis of this data set serves two purposes. First, we want to compare post selection inference results with an existing method. Second, we want to verify that minimizing \eqref{minpen2c} with real data can provide a sensible graph of the responses. For the latter issue the answer is yes. Table \ref{genomicsA} provides the entries of $\hat{\vA}$ and aligning with the scientific literature the optimal penalty is to penalize a difference in the weight measurements and a difference in the cotinine measurements, while all other comparisons the penalty enforces a negative correlation between the measurements. 

\begin{table}[ht]
\centering
\resizebox{.6\textheight}{!}{\begin{tabular}{rrrrr}
\hline
& Placental Weight & Newborn Weight & Blood Cotinine & Cord Cotinine \\
\hline
Placental Weight & 0 & -1 & 1 & 1 \\
 Newborn Weight& -1 & 0 & 1 & 1 \\
Blood Cotinine & 1 & 1 & 0 & -1 \\
 Cord Cotinine & 1 & 1 & -1 & 0 \\
\hline
\end{tabular}
}
\caption{ The $\vA$ matrix for the genomics data}
\label{genomicsA}
\end{table}

Figure \ref{fig:allTable} provides the post selection confidence intervals for the lasso and MinPen methods. The selected models are very similar with the biggest model difference being in the model of the weight for placenta, where MinPen selects four more variables include the mother's age which is the only demographic variable difference between the models. In addition, for this model the post selection results for MinPen find a negative relationship between smoking and placenta weight, but the lasso results are inconclusive. The previously cited literature appears indicates that for this particular relationship the lasso approach suffers from a Type II error, while MinPen correctly identifies this relationship. However, for modeling cotinine level in the umbilical cord, the one setting where the variables selected are the same, the lasso confidence intervals are noticeably smaller.

\begin{figure}
	\centering
		\includegraphics[width=0.8\textwidth]{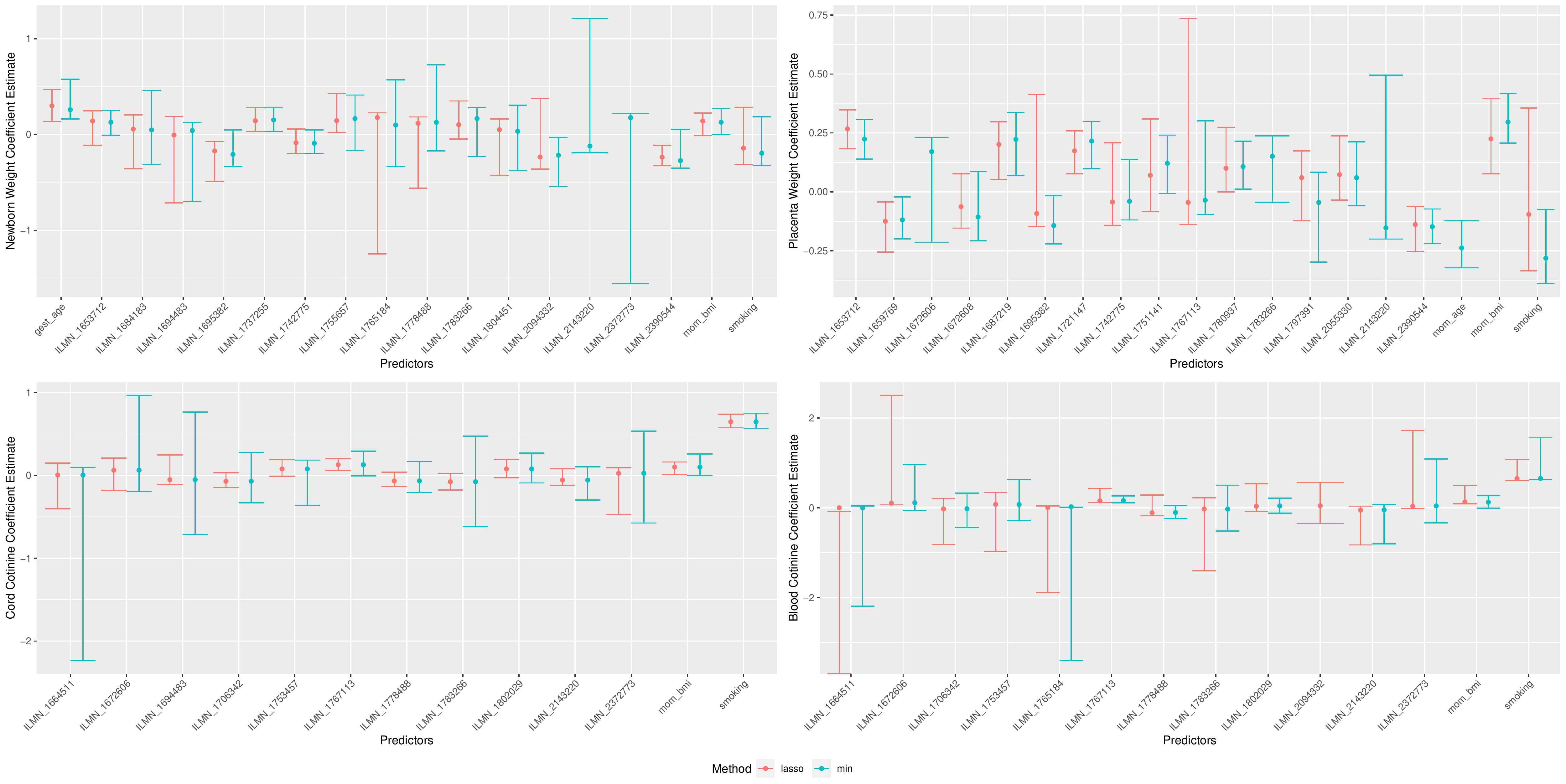}
	\caption{Post Selection Confidence Intervals: Clockwise starting in the top left corner; (1) Newborn weight; (2) Placenta Weight; (3) Umbilical cord cotinine level; (4) Peripheral blood cotinine level }
	\label{fig:allTable}
\end{figure}

\subsection{Connecticut Multiple Substance Use Data}
\label{SubstanceUse}

We investigate the use of the minimum penalty to identify co-occurring drug use based on accidental drug related deaths in Connecticut from 2012-2018.  The data set consists of 5097 accidental overdoses in the state of Connecticut and corresponding toxicology, death certificate, and scene investigation information from the Chief Medical Examiner Office,  publicly available at https://data.ct.gov and included in the supplemental material.  We identify 17 response variables which are 16 individual drugs that were investigated plus a response that identifies if any opiate is involved.  The goal of this analysis is to identify how different drugs appear (or do not appear) in cases of overdose, with a hope of better understanding and being able to predict co-occurrence of drugs.  By being able to predict co-occurrence of drugs, both clinical practitioners and law enforcement are better able to understand emerging trends in drug usage allowing them to quickly identify, intervene, and mitigate issues.  

The covariates in this data are indicators of the city and state of the residence and death, along with using a bag of words approach to find common language used in the cause of death and description of injury provided by police and the medical examiner.  As the goal is to better understand the relationships between drugs being used, information in the cause of death may not be relevant to all drugs found in the system.  In total 111 covariates were used in this analysis.  A validation set was created from 500 randomly selected observations from 2018  tuning parameters were selected using a 90/10 train/test approach using total classification error rate.  Once the tuning parameters were obtained the model was fit using the chosen tuning parameters on all data less the validation set.  We fit MinPen, MCEN, and elastic net models using a common sparsity parameter for all 17 responses, and compare methods using ROC curves on the validation set, see Figure \ref{abuse_roc}.  The results show the MinPen is competitive or out performs MCEN and the elastic net approach.  Furthermore,  MinPen provides relationships between responses.  Only positive or no relationships were found between responses, that is no negative relationships were found. The supplementary material includes a graph of these relationships.  Most notably heroin is not predictive or related to fentanyl,  fentanyl analogue,  benzodiazapine, or inform the presence of other opioids.    The opioid variable was mostly related to prescription based drugs such as oxymorphone,  hydrocodone, methadone, among others.  Methadone was positively related to every other variable which is notable because it is used as a medical assisted treatment for opiate abuse.   Further, the model finds benzodiazapine is not related to cocaine, which is of interest as benzodiasipine is used to treat cocaine toxicity. % \bfblue{(If we want to include the graph in the supplemental material then I think we should inlclude the discussion of the relationships there as well.)} \bfred{Not sure about that here, my biggest concern is that it's of interest but we're not reporting.  If we have room I say leave it if not cut it and then go to supplement.}    MCEN produced 2 clusters,  a small cluster with morphine, fentanyl analogue, hydromorphine, and OpiateNOS (not on scale), and a larger cluster with all other responses.  
The results show MinPen provides better prediction accuracy than both MCEN and elastic net, providing evidence that use of the minimum penalty can be beneficial in the presence of complex and potentially unknown relationships.   In the supplemental material we also present a heatmap of the coefficients produced by MinPen for each response.% showing the non-zero variables in the model.  

\begin{figure}
\begin{center}
\includegraphics[height=3in, width=4in]{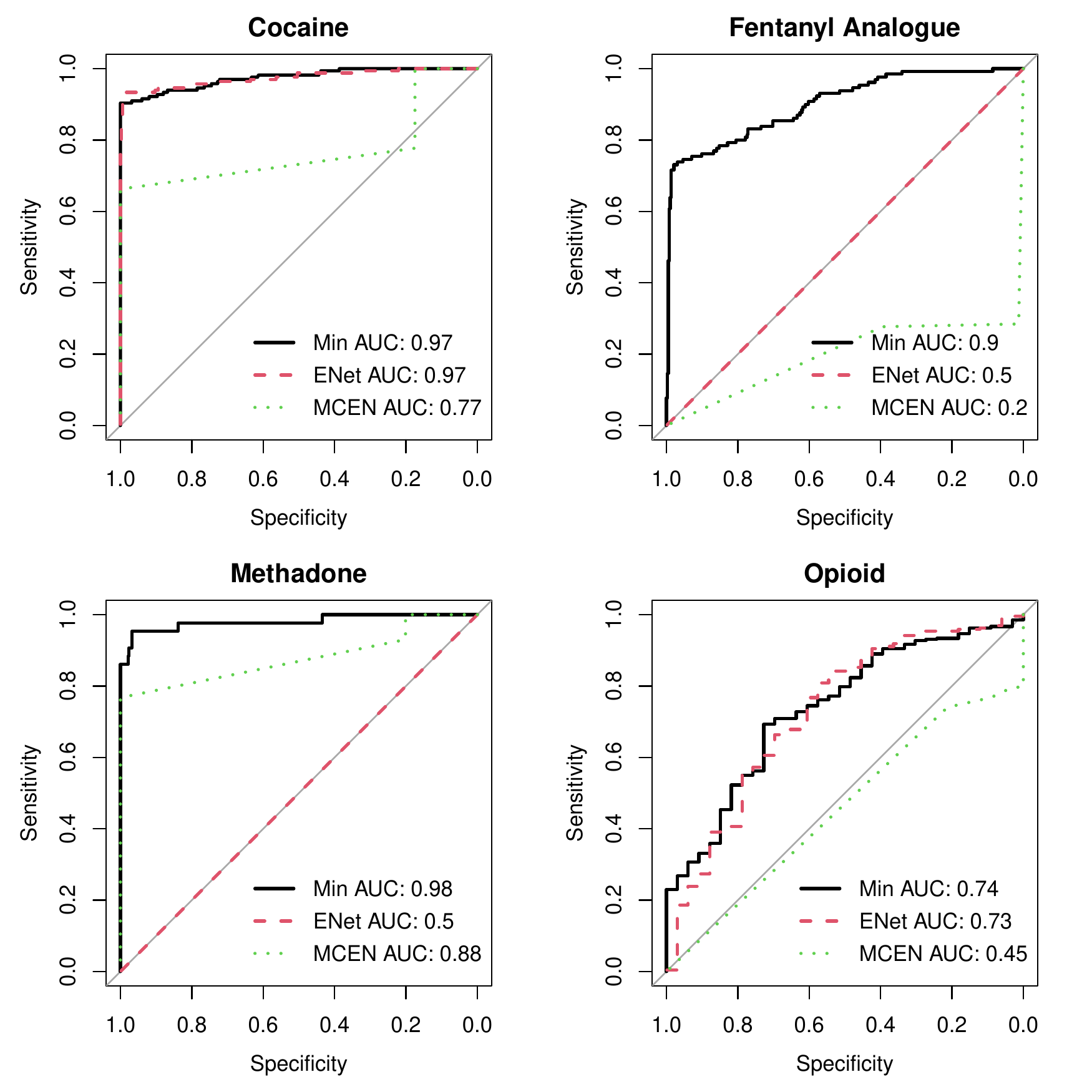}
\end{center}
\caption{Results of the validation set of 500 observations from 2018 with regard to four substances:  Cocaine,  Fentanyl Analogue,  Methadone,  and the ability to detect any opioid.  The results show that MinPen performs as well if not better with regard to AUC in classification.  }
\label{abuse_roc}
\end{figure}

%\begin{figure}
	%\centering
		%\includegraphics{plTable.pdf}
	%\label{fig:plTable}
%\end{figure}
%
%
%\begin{figure}
	%\centering
		%\includegraphics{newbornTbl.pdf}
	%\label{fig:newbornTbl}
%\end{figure}
%
%
%
%\begin{figure}
	%\centering
		%\includegraphics{cordTable.pdf}
	%\label{fig:cordTable}
%\end{figure}
%
%
%\begin{figure}
	%\centering
		%\includegraphics{bloodTable.pdf}
	%\label{fig:bloodTable}
%\end{figure}

\bibliographystyle{agsm}

\bibliography{min_pen}

\end{document}